\documentclass[a4paper,fleqn]{cas-sc}

\usepackage[numbers]{natbib}
\usepackage{subcaption}
\usepackage{siunitx}
\DeclareSIUnit\angstrom{\text{Å}}
\usepackage{placeins}
\usepackage[version=4]{mhchem}
\usepackage{booktabs}
\usepackage{multirow}
\usepackage{bookmark}
\usepackage{hyperref}

\begin{document}
\let\WriteBookmarks\relax
\def\floatpagepagefraction{1}
\def\textpagefraction{.001}

\shorttitle{In the age of big data and AI: do we still need chemical bonds?}

\shortauthors{Al-Fahdi et al.}

\title [mode = title]{In the age of big data and AI: do we still need chemical bonds?}                      


%
\author[1]{Mohammed Al-Fahdi}[type=editor,
                        auid=000,
                        orcid=0000-0001-5296-8433]

\credit{Conceptualization, Methodology, Software, Validation, Formal analysis, 
Investigation, Data curation, Writing – Original Draft, Writing $-$ Review \& Editing, 
Visualization}


\author[1,2]{Aakash Naik}[type=editor, orcid=0000-0002-6071-6786]

\credit{Methodology, Software, Validation, Formal analysis, 
Investigation, Data curation, Writing – Original Draft, Writing $-$ Review \& Editing, 
Visualization}

\author[1,2]{Katharina Ueltzen}[type=editor, orcid=0009-0003-2967-1182]

\credit{Methodology, Software, Validation, Formal analysis, 
Investigation, Data curation, Writing – Original Draft, Writing $-$ Review \& Editing, 
Visualization}

\author[1,2]{Luc Walterbos}[type=editor, orcid=0009-0000-5992-4353]

\credit{Methodology, Software, Formal analysis, 
Investigation, Writing – Original Draft, Writing $-$ Review \& Editing, 
Visualization}

\author[1]{Tom Demeyere}[type=editor, orcid=0000-0002-5023-6156]

\credit{Software, Validation, Writing – Original Draft, Writing $-$ Review \& Editing, 
Visualization}

\author[1]{Philipp Benner}[type=editor, orcid=0000-0002-0912-8137]

\credit{Conceptualization, Methodology, Software, Validation, Formal analysis, 
Investigation, Writing – Original Draft, Writing $-$ Review \& Editing, 
Visualization, Supervision}


\author[1,2]{Janine George}[type=editor, orcid=0000-0001-8907-0336]

\cormark[1]

\ead{janine.george@bam.de}

\credit{Conceptualization, Methodology, Software, Validation, Formal analysis, 
Investigation, Data curation, Writing – Original Draft, Writing $-$ Review \& Editing, 
Visualization, Supervision, Project administration, Funding acquisition}

\affiliation[1]{organization={Bundesanstalt für Materialforschung und -prüfung (BAM)},
    addressline={Unter den Eichen 87}, 
    city={Berlin},
    postcode={12205}, 
    country={Germany}}

\affiliation[2]{organization={Friedrich Schiller University Jena},
    addressline={Max-Wien-Platz}, 
    city={Jena},
    postcode={07743}, 
    country={Germany}}

\cortext[cor1]{Corresponding author}

\begin{abstract}
Chemical bonding is the language through which chemistry rationalizes materials structure, stability, properties, and reactivity. Meanwhile, modern deep-learning models achieve remarkable predictive performance using representations based largely on compositional and structural information, raising the question of whether explicit bonding concepts remain necessary. In this work, we argue that artificial intelligence does not render chemical bonding obsolete but provides a new framework for evaluating and refining bonding concepts. We show that, through linear probing, Allegro and MACE embeddings recover bonding descriptors with ($R^2$ $>$ 0.95), indicating that bonding emerges naturally from Allegro's and MACE's latent spaces. We also show that bonding descriptors provide information beyond bond distance in our electronic-structure-featurized MEGNet (ESF-MEGNet) model, reducing prediction error for properties including bulk and shear moduli by 15\%. We discuss implications for interpretable machine learning, synthesizability prediction, and generative materials design, and suggest joint-embedding approaches for learning more general bonding representations. More broadly, we propose that chemical bonding can be understood as a learnable representation of electronic structure. From this perspective, different bonding models represent alternative compressions of the underlying electronic information and can be evaluated through their predictive power, dimensionality, and complexity. The age of machine learning shifts the focus from whether a bonding model is physically motivated to whether it captures the information required to explain and predict materials behavior. If a materials genome exists, it should be encodable through a representation of chemical bonding. If no bonding representation can capture this information, our understanding of the chemical bond may require revision.

\end{abstract}

\begin{keywords}
Chemical Bonding \sep Deep learning \sep Materials Chemistry \sep Trustworthy AI
\end{keywords}

\begin{NoHyper}
\maketitle
\end{NoHyper}

\section{Introduction}

One of the most fundamental heuristics in chemistry is the chemical bond heuristic.\cite{pauling_nature_1960} Yet chemical bonds are not quantum-mechanical observables and therefore cannot be defined uniquely from quantum mechanics. For many organic molecules, this ambiguity is often of limited practical consequence: structures composed mainly of carbon, hydrogen, light main-group elements, and halogens can frequently be rationalized using familiar bonding concepts, ranging from the Lewis picture to more formal interpretations based on molecular-orbital theory. For example, the Woodward-Hoffmann rules have successfully predicted chemical reactivity.\cite{woodward_stereochemistry_1965} In inorganic materials and extended solids, however, the situation is considerably less clear-cut, despite existing analogies and relationships to (organic) molecules\cite{hoffmann_building_1982, hoffmann_how_1987} and many success stories. \cite{burdett_chemical_1995} A much larger region of the periodic table becomes relevant, different bonding motifs may coexist, and the chemical bonding heuristic has not achieved the same generality and practical success as in molecular chemistry.\cite{Miller2017-kc}

The ambiguity of the chemical bonding definition has led to a wide range of chemical bonding descriptors, none of which can provide a uniquely correct definition of the chemical bond by construction. Alongside quantum-chemical approaches such as orbital- and electron-density-based methods, researchers have also characterized bonding through geometrical descriptors based on interatomic distances and coordination environments. Orbital-based descriptors play a central role in chemical reasoning, partly because molecular orbital theory is introduced early in chemical education. Beyond their pedagogical importance, molecular orbitals also provide an intuitive framework for rationalizing bonding mechanisms.\cite{pan_critical_2021} In contrast, electron-density-based descriptors connect more naturally to physical and crystallographic reasoning, since electron density is directly related to observable quantities obtained, for example, from diffraction experiments. While the question of bonding has been debated from different conceptual perspectives, it has also been approached pragmatically. If no uniquely correct definition of a chemical bond exists, the relevant question is not which descriptor is "true", but which descriptor is most useful for a given scientific task, whether prediction or chemical reasoning. As Dewar famously remarked, "The only criterion of a model is usefulness, not 'truth'."\cite{dewar_chemical_2002, pan_critical_2021}

Machine learning (ML) further reinforces this pragmatic perspective. Evaluating such usefulness, however, has traditionally relied on chemical intuition, qualitative arguments, or selected case studies. In the era of data-driven science, researchers can assess the usefulness of bonding descriptors more quantitatively than before. Machine-learning methods provide a framework for evaluating bonding descriptors through their correlations or mutual information with materials properties and reactivities, as well as their impact on the predictive accuracy, robustness, transferability, and interpretability of ML models. From this perspective, a bonding descriptor's value is not determined by its success in explaining selected examples of chemical bonding, but by how well it captures chemically relevant information for a given prediction task. We propose that a useful bonding descriptor might be defined as a low-dimensional compact representation of the electronic structure that shares high information content with a target property, can ultimately be connected to chemical concepts and heuristics, and allows chemical reasoning.

\begin{figure}[pos=h]
    \centering
    \includegraphics[width=0.7\linewidth]{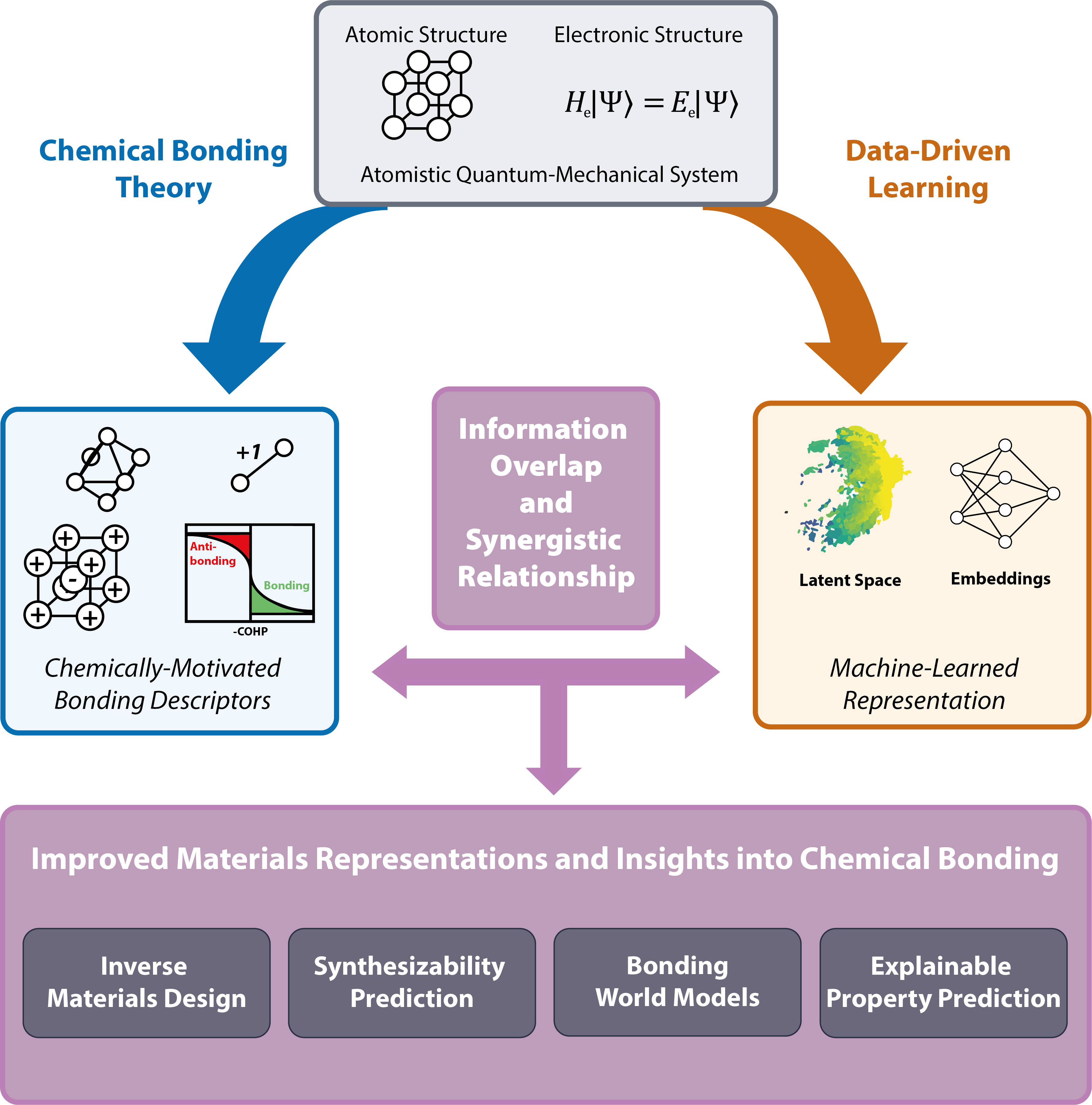}
    \caption{Conceptual overview of the relationship between chemical bonding and machine learning. This work explores how data-driven methods can assess the utility of bonding descriptors, whether modern graph neural networks implicitly learn chemically meaningful bonding information, and how bonding concepts can improve the interpretability, performance, and ultimately trust in machine-learning models.}
    \label{fig:overview}
\end{figure}

Deep learning places the concept of chemical bonding under renewed pressure. Modern architectures that rely primarily on atomic identities and interatomic distances have been shown, for example, to reproduce energies, forces, and stresses with remarkable accuracy. Once sufficiently large datasets are available, explicit bonding concepts may appear unnecessary for achieving accurate property predictions. In a similar spirit, one may ask whether explicit physical constraints, such as symmetry equivariance, are strictly necessary if models are sufficiently flexible and data are abundant.\cite{PET-ECSE-2023,pushing-unconstrained-2026}

However, such constraints could still be valuable because they encode physically and chemically meaningful inductive biases, particularly in scenarios with limited data. Discarding the concept of a chemical bond in chemistry could be seen as similarly controversial as discarding the concept of symmetry in physics. Deep learning may provide an opportunity to revisit chemical bonding from a different perspective. Models trained only on atomic structures and properties may nevertheless learn internal latent representations that encode information related to chemical bonding.\cite{hattori2026chemical} The benefit of interpreting these representations in terms of chemical bonding concepts could be twofold: (1) We might rediscover known bonding descriptors and concepts, thereby increasing trust in deep-learning models for materials property prediction. (2) We might identify bonding descriptors more closely connected to, and predictive of, materials properties and reactivity. 

In this work, we explore the role of chemical bonding concepts in an era increasingly dominated by deep learning. We argue that the question of how chemical bonding should best be described can benefit from a data-driven perspective focused on the utility of bonding descriptors for predicting and understanding materials properties. Modern machine-learning approaches provide new opportunities to quantitatively evaluate, compare, and exploit bonding concepts. We review the motivation for incorporating quantum-chemical bonding descriptors into crystal-structure representations, discuss the ambiguity of quantum-chemical bonding descriptors and the current state of data-driven bonding analysis, and examine recent advances in deep learning, including graph neural networks and machine-learned interatomic potentials. We then show how deep-learning models for crystal potential energy surfaces (machine-learned interatomic potentials, MLIPs) implicitly encode bonding information and introduce a graph-based machine-learning architecture that incorporates quantum-chemical bonding descriptors and atomic charges, improving both property prediction and model interpretability. Finally, we discuss opportunities for bonding-aware machine learning, including sparse-data applications, generative models, and synthesizability prediction; the advantages of joint-embedding predictive architectures for combining multi-modal bonding information; and foundation models for interatomic potentials and bonding analysis. This work can be viewed as a sequel to Ref.~\cite{george_chemist_2021}, which discussed the interplay between chemical heuristics and machine learning. While that work considered chemical heuristics more broadly, this work focuses on chemical bonding in the era of data-driven materials informatics and its relationship to modern machine-learning models. (Fig.~\ref{fig:overview}).

\section{Chemical Bonding Descriptors as Representations of Electronic and Atomic Structure}

A wide range of descriptors has been developed to rationalize crystal structures and materials properties. In this work, we view these approaches as representations of the underlying atomic and electronic structure that differ in the type and amount of information retained. For example, coordination environments,\cite{pan_benchmarking_2021,waroquiers_chemenv_2020} bond-valence models,\cite{brown_bond_1978} lattice-energy approaches,\cite{hoppe_madelung_1995} and modern orbital-, density-, and energy-based descriptors\cite{hughbanks1983chains,dronskowski1993crystal,mueller2021crystal,bader1990quantum,kreuter2025energy} therefore represent different levels of abstraction of the same quantum-mechanical system. Rather than fundamentally different physical pictures, they emphasize different aspects of the atomic and electronic structure and therefore of chemical bonding. They also differ in their compactness, that is, in how efficiently they compress electronically relevant information into chemically interpretable quantities. This viewpoint is particularly relevant in machine learning, where a representation's usefulness depends on its ability to predict and potentially explain material properties.

Among the earliest and still most influential descriptors are coordination environments. Introduced into crystal chemistry by Pfeiffer and later incorporated into Pauling's crystal-chemical rules,\cite{pfeiffer_kristalle_1915,pfeiffer_krystalle_1916,pauling_principles_1929} they relate local atomic arrangements to structural stability and continue to shape the interpretation of inorganic crystal structures.\cite{george_limited_2020} Coordination environments identify local atomic interactions but do not explicitly distinguish between different bonding mechanisms. Nevertheless, they provide a compact and highly effective description of local structure and its relationship to materials properties.

Extending these concepts, bond-valence theory introduced a more quantitative description of local interactions.\cite{brown_bond_1978} By assigning bond strengths through empirical bond-length relationships, it provides a simple framework for rationalizing structural stability, distortions, and aspects of chemical reactivity. Similarly, electrostatic lattice-energy models such as Hoppe's MAPLE approach use formal oxidation states and crystallographic structure information to estimate lattice energies of ionic solids.\cite{hoppe_madelung_1995} Both provide compact, low-dimensional chemically useful representations of atomic and electronic structure.

The development of quantum mechanics shifted attention from geometric and electrostatic models toward electronic theories of bonding. Valence-bond (VB) and molecular-orbital (MO) theory established the foundations of modern quantum chemistry and inspired bonding descriptors derived directly from electronic-structure calculations.\cite{pan_critical_2021} In solids, these include orbital-based approaches such as Crystal Orbital Overlap Populations (COOP), Crystal Orbital Hamilton Populations (COHP), and Crystal Orbital Bond Indices (COBI), density-based methods such as the Quantum Theory of Atoms in Molecules (QTAIM), and energy-partitioning approaches.\cite{hughbanks1983chains,dronskowski1993crystal,mueller2021crystal,bader1990quantum,kreuter2025energy} While these approaches differ substantially in their underlying philosophy and interpretation, they all seek to extract chemically meaningful information from the electronic structure. The physical meaning and interpretation of orbital-, density-, and energy-based bonding descriptors remain the subject of ongoing debate.\cite{pan_critical_2021}

Despite their conceptual differences, bonding descriptors are increasingly employed as features for both materials properties and discovery.\cite{oestereich_decoding_2024,Ram2025-iv,Naik_Dhamrait_Ueltzen_Ertural_Benner_Rignanese_George_2026} This use is perhaps unsurprising, as many of these descriptors were originally empirically developed and validated based on their ability to organize, rationalize, and predict experimental observations. For example, numerous empirical models predicting mechanical and thermal properties rely on bonding-related quantities such as coordination environments, bond lengths, ionicity, bond strength, bond overlap populations, and valence-electron concentrations.\cite{Gilman_1975,Phillips_1970,Jhi_Ihm_Louie_Cohen_1999,Garbato_Rucci_1979,Guo_Li_Liu_Yu_He_Liu_Xu_Tian_Wang_2008,Muttalib_Abdullah_2025,Hassan_Omar_2025,holzer_silicate_2023}

This representation-based viewpoint underpins the machine-learning perspective developed in the following sections. If bonding descriptors are viewed as compact representations of electronic structure, it is natural to ask whether machine-learning models can discover alternative representations that retain similar or even greater amounts of chemically relevant information. Deep learning based on modern graph neural networks provides perhaps the most successful example of such learned representations in materials science, as these models construct latent descriptions of atomic environments directly from data. The following sections therefore explore the relationship between these learned representations and established concepts of chemical bonding.

\section{Deep Learning of Potential Energy Surfaces and Materials Properties}\label{sec:DL_PES_MatProp}

Machine-learned interatomic potentials (MLIPs) and other graph-based models have been developed to learn potential-energy surfaces and materials properties.\cite{chen_graph_2019,chen_universal_2022} Today, these approaches are predominantly deep learning-based and have achieved remarkable success in accurately predicting a wide range of materials properties. Pretrained models have emerged that, much like large language models, can be fine-tuned for related datasets or even repurposed entirely from one materials-property prediction task to another.\cite{batatia_foundation_2025, riebesell2025framework,grandel_targeted_2026,tompa_fine-tuning_2026,xiao_accurate_2026} Remarkably, different architectures seem to converge toward similar underlying representations.\cite{Li_Walsh_2026} The latent representations also help in generating new stable materials.\cite{hagemann_representations_2026}

The prevailing assumption is often that more data will lead to better models. While this may hold for tasks such as learning ab initio energies, forces, and stresses, where large training datasets can be generated systematically, many other problems in materials science are ultimately constrained by limited data availability. In such settings, particularly for experimental datasets, models frequently encounter sparsely sampled regions of the input space where extrapolation becomes essential.\cite{li_probing_2025} Transfer learning offers one strategy to address this challenge, but alternative approaches, including improved representations and the integration of complementary information from multiple data modalities, may also improve generalization beyond the available data.

Deep-learning architectures can appear surprisingly complex from a chemical perspective. However, many ultimately rely on mathematical descriptions of local atomic environments, incorporating information from large numbers of neighboring atoms through mechanisms such as message passing to learn extensive or intensive properties.\cite{chen_universal_2022, batatia_mace_2022} In machine-learned potential energy surfaces, these local environment representations are typically used to predict atomic energies, which are then summed to obtain the total energy. Although the resulting latent representations are rarely expressed in the language of established chemical concepts and heuristics, the graph-based representations themselves already imply a notion of local atomic interactions that bears a strong resemblance to traditional concepts of chemical bonding. This naturally raises the question of whether the latent representations learned by such models encode information that chemists would associate with chemical bonding.  One architecture that makes this connection between deep-learning models and chemical bonding more explicit is Allegro, where bond energies are combined to produce the final energy prediction.\cite{musaelian_learning_2023} 

Both atomic and bond energies derived from such models have already been used to rationalize materials properties in chemically meaningful ways, suggesting a connection between the learned representations and traditional concepts of chemical bonding.\cite{wu_zintlklemm_2026,hattori2026chemical} 

These observations raise a fundamental question: what are these models actually learning? And, further, are modern graph neural networks implicitly recovering traditional bonding concepts, or are they learning distinct representations that only partially overlap with established chemical descriptors? A third possibility is that both representations ultimately contain similar information, but that incorporating chemically motivated bonding descriptors makes learning more data-efficient, robust, and interpretable. In this view, bonding descriptors do not necessarily introduce new information; rather, they provide a useful inductive bias that helps models learn relevant chemical relationships from limited data, analogous to the role of symmetry information in modern machine-learning models. Exploring these connections may not only improve our understanding of modern deep-learning architectures but also make them more accessible to chemists by relating their learned representations to familiar chemical concepts and heuristics. We will explore these questions further within Sec.\,\ref{sec:mlips_bondinginfo}. Before doing so, however, we return to the bonding descriptors themselves and discuss the ambiguities inherent in their definition.

\section{Ambiguity in Bonding Representations and Algorithms}\label{sec:bonding_amb}

If bonding descriptors are viewed as alternative representations of the atomic and electronic structure, an immediate question is how different representations compare in practice. We compare bonding models derived from electronic-structure calculations with geometric descriptions of crystal structures. In the latter, bonds are typically assigned using interatomic distances or angles, often supplemented by empirical concepts such as atomic, ionic, or covalent radii. While these approaches are computationally inexpensive and successful for many classes of materials, they may not capture all chemically relevant interactions in compounds with complex or ambiguous bonding, such as intermetallic compounds.\cite{pan_benchmarking_2021} Quantum-chemical bonding descriptors provide access to electronic-structure information, but they are computationally more demanding and may lead to different, rather than uniquely correct, bond assignments. 

The \ce{ThCr2Si2} structure type provides an instructive example in which geometric and quantum-chemical representations succeed and fail in complementary ways. Importantly, the ambiguity arises not only from the underlying bonding representation but also from the algorithms and criteria used to convert these representations into discrete bond assignments.

\begin{figure}
    \centering
    \includegraphics[width=0.9\linewidth]{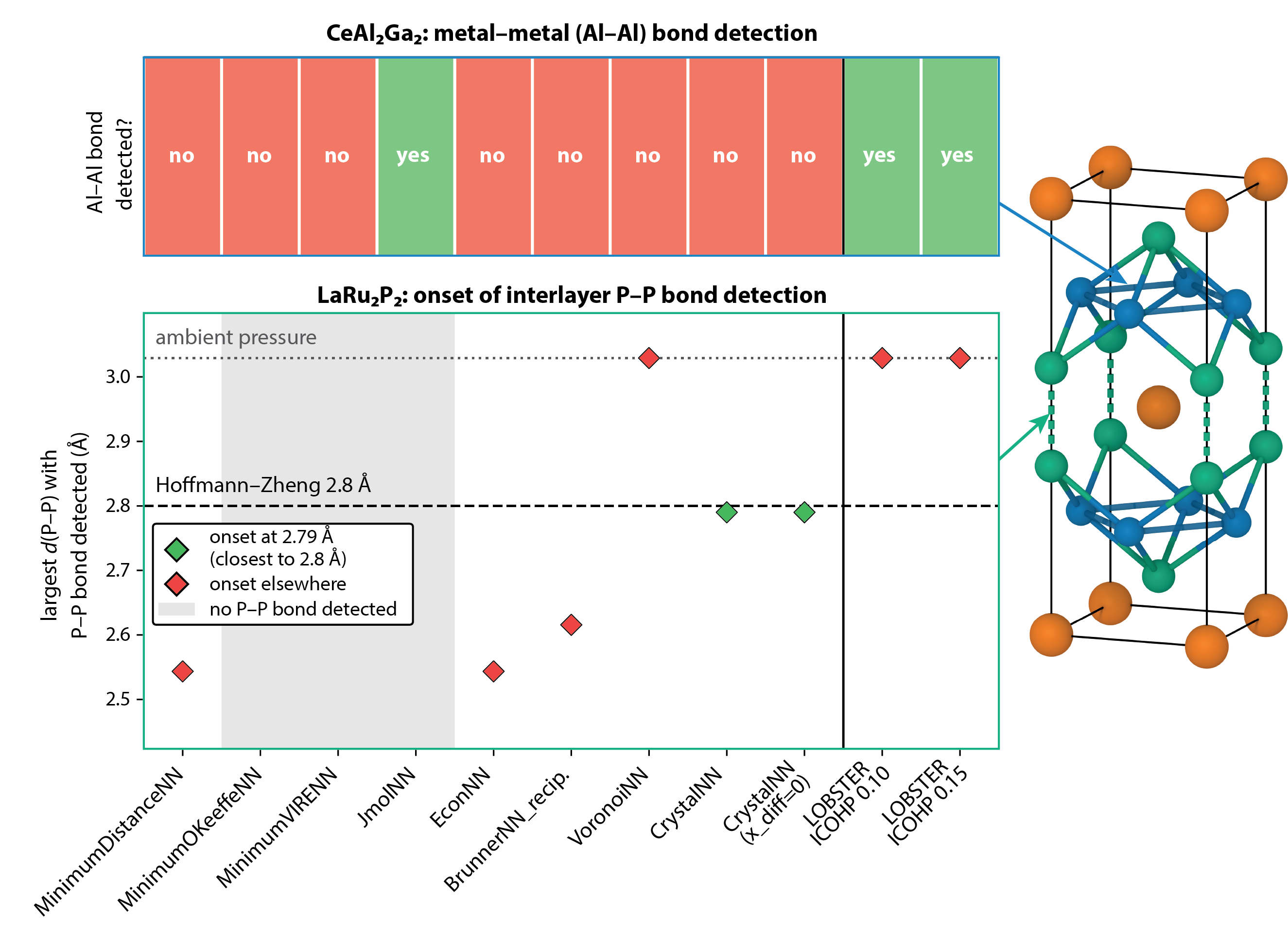}
    \caption{The \ce{ThCr2Si2} structure type (right) and the performance of geometry-based vs. quantum chemical bonding-based (LOBSTER\cite{deringer2011crystal, maintz2013analytic, maintz2016lobster,nelsonLOBSTERLocalOrbital2020}) coordination detection algorithms (left). The geometry-based algorithms are documented in detail in \cite{pan_benchmarking_2021}. CrystalNN(x\_diff=0) omits the electronegativity weighting of the CrystalNN algorithm, which was previously found to improve the description of intermetallic bonding.\cite{pan_benchmarking_2021} All but one geometry-based algorithm fail to detect metal--metal bonding, while LOBSTER correctly recognizes it. However, LOBSTER does not predict the onset of P--P bonding in \ce{LaRu2P2} at \SI{2.8}{\angstrom} as proposed by Hoffmann and Zheng.\cite{hoffmann_making_1985} A geometry-based algorithm, CrystalNN,\cite{pan_benchmarking_2021} correctly captures the onset of interlayer P--P bonding in this compound.\cite{foroozani2014}}
    \label{fig:thcr2si2-type}
\end{figure}

Compounds of the \ce{ThCr2Si2} type share the composition \ce{AB2X2}, where B = transition metal and X = a group 14 or 15 element (P, As, Si, Ge,...). They crystallize in $I$4/$mmm$ with A on the 2$a$ site (0, 0, 0), B on 4$d$ (0, 0.5, 0.25), and X on 4$e$ (0, 0, z) (see Fig.\,\ref{fig:thcr2si2-type}).
Multiple types of bonding coexist in this structure type: ionic bonds between the A cation and \ce{B2X2} anionic layers, B--X covalent bonds, B--B metal-metal bonds, and, in some compounds of this structure type, interlayer X--X bonds.\cite{hoffmann_making_1985, shatruk_thcr2si2_2019}
The properties are strongly affected by the different types of bonding in these materials: small variations in the metal-- metal bonds influence the magnetic structures of many \ce{ThCr2Si2}-type compounds.
Superconductivity in several \ce{ThCr2Si2}-type compounds has been attributed to the absence of X--X bonding.\cite{shatruk_thcr2si2_2019}

One compound of this structure type is part of a benchmark for purely geometric bonding-detection algorithms in crystal structures.\cite{pan_benchmarking_2021}  Overall, these algorithms reproduce human-assigned coordination numbers well for elemental or highly ionic structures. However, they perform significantly worse for intermetallic compounds.  Interestingly, all but one benchmarked geometric bonding analyses fail to detect metal-metal bonding in several compounds of the \ce{ThCr2Si2} structure type. Consequently, a high-throughput screening campaign for metal-metal bonds using geometry-based algorithms would overlook them in this structure type.

This limitation of the geometric bonding algorithms can be overcome by using bond strengths derived from quantum-chemical bonding analyses (Integrated Crystal Orbital Hamilton Populations, ICOHPs)\cite{dronskowski1993crystal} to detect coordination instead: In this algorithm, two sites are considered bonded if their ICOHP is at least 10\,\% of the strongest ICOHP in the structure. With this approach, metal-metal bonds are correctly determined in the tested \ce{ThCr2Si2}-type compounds.

However, the ICOHP-based coordination criterion performs less well for another important bonding motif in \ce{ThCr2Si2}-type compounds. Several \ce{ThCr2Si2}-type compounds with X=P show superconductivity, which has been attributed to the absence of P--P bonding.\cite{shatruk_thcr2si2_2019} Hoffmann and Zheng determined a P--P distance of about 2.8\,\AA{} as the threshold for covalent bond formation in this structure type.\cite{hoffmann_making_1985}
For example, the superconducting \ce{LaRu2P2}, which has a P--P distance of around 3.0\,\AA{} at ambient conditions, loses its superconductivity at pressures of 2.1\,GPa, which was attributed to P--P bond formation.\cite{foroozani2014}

To investigate this behavior with different bonding algorithms,  we simulated the pressure onset with a series of DFT relaxations at reduced volumes and performed both geometric and ICOHP-based bonding analyses on the resulting structures.
The geometry-based CrystalNN\cite{pan_benchmarking_2021} correctly locates the onset of P--P bonding near the 2.8\,\AA{} threshold, whereas the 10\,\%-ICOHP criterion assigns a relevant P--P bond at all volumes considered, including the ambient-pressure structure with a P--P distance of around 3.0\,\AA{}.

The energy-resolved COHP curves are nevertheless consistent with the model proposed by Hoffmann and Zheng\cite{hoffmann_making_1985}: as the P--P distance decreases, antibonding P--P states located near the Fermi level at ambient pressure shift above it, stabilizing the interaction. 
Although this qualitative trend is present in the COHP, a simple and somewhat arbitrary cutoff on the integrated COHP does not capture the expected onset of bonding. Additionally, higher-level DFT methods, e.g., different exchange-correlation functionals, might influence the results. 

To summarize, both purely geometric and current quantum-chemical bond-detection algorithms can fail to detect chemically relevant bonds in ambiguous bonding scenarios. In the \ce{ThCr2Si2} structure type, geometric algorithms fail to detect metal-metal bonding, whereas an algorithm based on quantum-chemical bond strengths overestimates X--X bonding. 

These observations highlight that different bonding representations and algorithms capture complementary aspects of the underlying electronic structure. Importantly, the failure of a particular bond-assignment algorithm does not necessarily imply that the underlying bonding representation is inadequate. Rather, it underscores that the conversion of continuous bonding descriptors into discrete bond assignments is itself a modeling choice. Data-driven approaches could be used to learn improved assignment criteria, for example by optimizing cutoffs or combining multiple descriptors against experimentally observed bonding-related phenomena.

Yet a more fundamental source of ambiguity exists. The examples discussed above arise from how different bonding representations and algorithms translate electronic structure into chemically interpretable bond assignments. However, even for a fixed electronic structure, different quantum-chemical partitioning schemes can yield different bonding descriptors. The origin of this non-uniqueness lies in the choices made when partitioning wave functions or electron densities into chemically meaningful quantities. We discuss these sources of ambiguity in the following section.

\section{The Non-Uniqueness of Quantum-Chemical Bonding Descriptors}\label{sec:limits_comp_mater_representation}

The ambiguity discussed in the previous section originated from different bonding representations and bond-assignment algorithms. However, even for a fixed electronic structure, chemically meaningful bonding descriptors are not unique. Bonding analyses require electrons to be assigned to interpretable labels, such as atoms, orbitals, or regions of space. None of these partitioning schemes is uniquely dictated by quantum mechanics, and their results depend on choices such as orbital localization, basis-set representation, projection procedures, or density-partitioning rules. Consequently, different bonding descriptors can yield different chemical pictures even when they originate from the same electronic structure.

In Hartree-Fock theory, the many-electron wave function is approximated by a single Slater determinant. Similarly, Kohn-Sham density functional theory (DFT) introduces an auxiliary Slater determinant constructed from occupied Kohn-Sham orbitals. In both cases, the occupied orbitals are only defined up to a unitary transformation among themselves. Equation~\ref{eq:gauge_freedom} shows such a transformation, which leaves all observables unchanged.

\begin{equation}
    \tilde{\psi}_{n} = \sum_{m} U_{mn} \psi_{m}
    \label{eq:gauge_freedom}
\end{equation}

While this freedom has no effect on observables, partitioning the electrons among atoms or bonds requires the occupied electronic states to be assigned to atomic, orbital, or bond contributions via a set of local reference functions. When these functions are built from the occupied states, typically including some unoccupied states, the choice of $\mathbf{U}$ directly affects such partitioning. This choice enters through the density matrix $\mathbf{P}$ in its basis representation, which, for an orthonormal basis set, is related to the total number of electrons $\mathcal{N}$ via Equation~\ref{eq:trace_dm}.

\begin{equation}
    \mathrm{Tr}(\mathbf{P}) = \mathcal{N}
    \label{eq:trace_dm}
\end{equation}

From $\mathbf{P}$, electrons can then be partitioned between bonds using the well-known Wiberg and Mayer bond-order frameworks,\cite{bridgemanMayerBondOrder2001} or Integrated Crystal Orbital Bond Index (ICOBI)-related quantities for periodic systems.\cite{mueller2021crystal} The resulting bonding picture therefore depends not only on the partitioning procedure but also on the basis used to represent the electronic states. Explicit orbital labeling within the basis representation is then desired to maximize the chemical interpretability of these quantities. However, such explicit labels are often in tension with computational accuracy. Equation \ref{eq:common_basis} displays two basis sets that are located at each extreme of chemical interpretability: 

\begin{equation}
    ^{\mathrm{PW}}\psi_{j}^{(\mathbf{k})}(\mathbf{r}) = \frac{1}{\sqrt{\Omega}}e^{i \mathbf{k}\cdot \mathbf{r}}\sum_{\mathbf{G}} C_{j}^{(\mathbf{k})}(\mathbf{G})e^{i \mathbf{G}\cdot \mathbf{r}} \qquad ^{\mathrm{STO}}\chi_{n,l,m}(\mathbf{r_{A}}) = N \, r_{A}^{n-1} \, e^{-\zeta r_{A}} \, Y_l^m(\theta_A,\phi_A)
    \label{eq:common_basis}
\end{equation}

where, for plane waves (PW), $\mathbf{k}$ is a wavevector in the Brillouin zone, $\mathbf{G}$ are reciprocal lattice vectors, $C_{j}^{(\mathbf{k})}(\mathbf{G})$ are the expansion coefficients of band $j$, and $\Omega$ is the cell volume; and, for Slater-type orbitals (STOs), $r_A = |\mathbf{r} - \mathbf{R}_A|$ is the distance to atom $A$ with polar angles $\theta_A$ and $\phi_A$, $n$, $l$ and $m$ are the principal, angular and magnetic quantum numbers, $\zeta$ is the orbital exponent, $Y_l^m$ is a spherical harmonic and $N$ is a normalization constant. Plane waves provide a complete and improvable basis set but cannot chemically label electronic states by atom or angular momentum. On the other side of the spectrum, STOs are atom-centered functions with well-defined angular momentum. They explicitly introduce interpretable labels, but typically form smaller bases, with reduced computational accuracy. Although extensions of these small bases are possible, for example, by adding polarization or diffuse functions, they further blur chemical interpretation, as with $d$ functions, which were read as $d$-orbital participation in hypervalent bonding,\cite{magnussonHypercoordinateMoleculesSecondrow1990} or Mulliken charges becoming unstable with diffuse functions.\cite{martinChargeDistributionWater2005} Therefore, the choice of basis set defines not only the span of the electronic states but also the chemical interpretability of such calculations, creating a direct and visible tension between accuracy and understanding.

In practice, bonding descriptors are often directly computed on a fixed basis set, which has been chosen to maximize chemical information. Doing so removes the conceptual dependence on $\mathbf{U}$ and ties the arbitrariness to the basis itself. In these cases, electronic minimization can directly be performed in the chosen basis set, as done in many popular codes such as FHI-aims for atom-centered numerical orbitals \cite{abbottRoadmapAdvancementsFHIaims2026} or GAMESS \cite{barcaRecentDevelopmentsGeneral2020} for Gaussian-type orbitals. Such minimization can also be first carried out in a complete, physically motivated basis set, e.g., plane waves using codes such as ABINIT \cite{verstraeteAbinit2025New2025} or Quantum ESPRESSO,\cite{giannozziAdvancedCapabilitiesMaterials2017a} and the resulting states can be projected onto a chemically interpretable basis set. This latter operation introduces the concept of spilling, which measures the deviation due to projecting states from a larger, more complete basis onto a smaller one.\cite{sanchez-portalProjectionPlanewaveCalculations1995} This is the approach taken by LOBSTER,\cite{deringer2011crystal, maintz2013analytic, maintz2016lobster,nelsonLOBSTERLocalOrbital2020} SSAdNDP,\cite{galeev2013solid} or PAOFLOW for tight-binding Hamiltonians.\cite{cerasoliAdvancedModelingMaterials2021} With these approaches, locality is imposed through the basis rather than through $\mathbf{U}$, even when basis functions are optimized for the system at hand, as in in situ optimized orbitals \cite{prenticeONETEPLinearscalingDensity2020} or spilling-minimized projection bases.

An alternative strategy is to recover interpretability directly through the unitary freedom represented by $\mathbf{U}$ but within the subspace of the Kohn-Sham states computed in a larger basis set, e.g., plane waves. That is, instead of projecting onto a different subspace, the new local functions are obtained by optimizing $\mathbf{U}$ for the desired properties, and bonding descriptors are computed on these rotated orbitals within the original subspace. Examples include selectively localized Wannier functions \cite{wangSelectivelyLocalizedWannier2014} and Pipek-Mezey Wannier functions.\cite{jonssonTheoryApplicationsGeneralized2017} However, the assumptions shift to the criterion that defines $\mathbf{U}$.

Instead of partitioning among basis functions, real-space bonding analysis requires assigning a region of space to atoms or bonds, often using topological or geometrical features. Because the electron density is invariant under orbital rotations, such approaches eliminate the dependence on the choice of $\mathbf{U}$ and largely decouple bonding analysis from basis-set effects. Expressing the same density-matrix operator $\hat{\rho}$, whose basis representation is $\mathbf{P}$, in the position basis gives the electron density $\rho(\mathbf{r})$ as its diagonal elements, as shown in Equation~\ref{eq:density_matrix_def}.

\begin{equation}
\rho(\mathbf{r}) = \rho(\mathbf{r},\mathbf{r}) = \langle \mathbf{r} | \hat{\rho} | \mathbf{r} \rangle = \sum_n f_n \, |\psi_n(\mathbf{r})|^2
\label{eq:density_matrix_def}
\end{equation}

In the position representation, the total number of electrons $\mathcal{N}$ is recovered via direct integration over space using Equation \ref{eq:electron_density_int}.

\begin{equation}
    \int \rho(\mathbf{r}) \; d\mathbf{r} = \mathcal{N}
    \label{eq:electron_density_int}
\end{equation}

However, the bonding picture still depends on how the density is partitioned. Partitioning the density requires specific rules such as zero-flux surfaces,\cite{baderAtomsMolecules1985} Voronoi cells,\cite{fonsecaguerraVoronoiDeformationDensity2004} or weighted schemes based on free-atom reference densities, such as Hirshfeld partitioning.\cite{hirshfeldBondedatomFragmentsDescribing1977} Again, results are sensitive to these choices, with these approaches yielding different atomic charges or bonding descriptors.

Both basis-based and real-space approaches rely on choices that are not uniquely dictated by the electronic structure, with different choices yielding different bonding descriptors for the same structure and composition. Consequently, the relationship between learned representations and chemical bonding cannot be assessed with respect to a unique bonding descriptor. Instead, any comparison is necessarily made relative to a particular bonding representation and its underlying assumptions.

Despite this non-uniqueness, bonding descriptors remain valuable because they compress complex electronic-structure information into forms that chemists can interpret and use for reasoning. Their non-uniqueness also suggests that bonding descriptors should be evaluated quantitatively rather than solely through chemical intuition or individual case studies. Notably, these descriptors are typically judged by their agreement with established chemical heuristics and their ability to rationalize observed behavior, whereas their compactness and effectiveness as representations of the underlying electronic structure are rarely assessed explicitly. Machine-learning models pose the complementary challenge: they learn highly predictive representations that are often difficult to rationalize chemically. This motivates the data-driven approach developed in the following sections, where we quantitatively evaluate bonding descriptors, explore their relationship to learned representations, and discuss whether multiple bonding analyses can be combined into more general representations of chemical bonding using approaches such as joint-embedding predictive architectures (JEPAs) (Sec.~\ref{sec:JEPA}).

\section{Toward a Data-Driven Science of Chemical Bonding}

\begin{figure}[pos=h]
    \centering
    \includegraphics[width=0.55\linewidth]{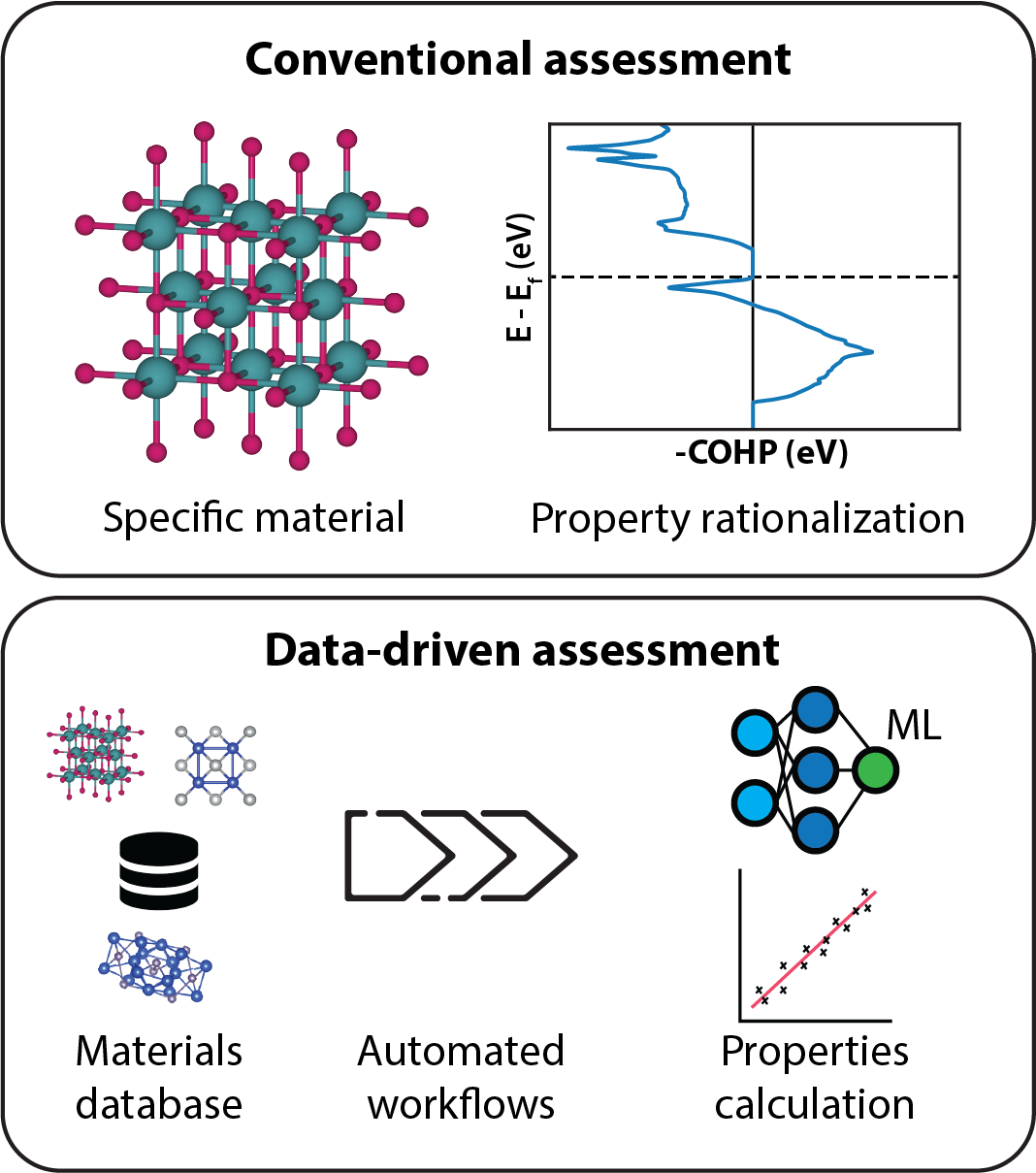}
    \caption{Evolution of chemical bonding from a rationalization framework to a quantitative assessment via statistical analysis and machine learning.}
    \label{fig:dd_chem_bonding}
\end{figure}

Viewing bonding descriptors as low-dimensional representations of electronic structure that support both prediction and reasoning naturally raises the question of how their usefulness should be evaluated. While these descriptors were originally developed to reason about materials properties, materials informatics and machine learning now provide a quantitative framework for assessing the amount of property-relevant information they contain across thousands of materials. This shift has transformed chemical bonding from a largely qualitative interpretive framework into a quantitative representation whose predictive utility can be analyzed statistically and exploited in machine-learning models (Fig. \ref{fig:dd_chem_bonding}). More generally, this perspective raises the possibility of evaluating bonding representations not only by predictive performance but also by how compactly they encode property-relevant information.

Among the various bonding representations proposed for solids, orbital-based descriptors derived from electronic-structure calculations are particularly attractive because they encode concepts familiar to chemists, such as bond orders, bond strengths, and bonding or antibonding interactions, while remaining directly connected to the underlying quantum-mechanical wavefunction. As a result, they support both property prediction and chemical reasoning, making them well suited for quantitatively assessing the usefulness of bonding representations in data-driven materials science.

The first prerequisite for such studies are computed bonding descriptors and therefore the ability to perform bonding analyses reliably and automatically across thousands of materials. Orbital-based bonding analysis using the software LOBSTER consists of two computational stages: a DFT calculation, followed by projecting the plane-wave wavefunction onto a localized, atomic-orbital-like basis to obtain a quantum-chemical description of chemical bonding. Each stage contains several subtasks, including input preparation, calculation execution, output validation, and post-processing. Performing such analyses for several thousand materials therefore requires robust automatic workflows.\cite{george2022automated}

To address this issue, automated workflows for LOBSTER calculations have been developed together with LobsterPy,\cite{naik2024lobsterpy} an open-source Python package for the automated analysis and interpretation of LOBSTER outputs. These workflows have been implemented within the open-source atomate\cite{mathew2017atomate,george2022automated} and atomate2\cite{ganose2025atomate2} Python packages, which provide necessary functionalities for high-throughput computational materials science and rely on several other community-developed Python packages.\cite{jain2015fireworks,ong2013python,george_pymatgen_2026,rosen2024jobflow,joblfowremote} Detailed descriptions of the workflow implementation are available in the corresponding publications. Similar high-throughput calculations have also been performed for other classes of bonding descriptors, enabling their integration into machine learning studies.\cite{Al_Fahdi_Rurali_Hu_Wolverton_Hu_2025, Okuda_Takahara_Mizoguchi_2026}

With the help of these automated workflows, some of us generated a publicly accessible bonding-analysis database comprising approximately 13,800\cite{Naik_Dhamrait_Ueltzen_Ertural_Benner_Rignanese_George_2026} crystalline materials obtained from the Materials Project.\cite{Jain_Ong_Hautier_Chen_Richards_Dacek_Cholia_Gunter_Skinner_Ceder_et_al_2013,horton2025accelerated} The database contains COOP, COHP, and two-center COBI curves together with their energy-integrated quantities (ICOOP, ICOHP, and ICOBI), as well as Mulliken and L{\"o}wdin atomic charges, projected densities of states (PDOS), and Madelung energies. The resulting database covers chemically diverse materials spanning the continuum between covalent and ionic bonding.\cite{Naik_Dhamrait_Ueltzen_Ertural_Benner_Rignanese_George_2026}

Using this database, previous work has systematically investigated the impact of LOBSTER-derived bonding descriptors in machine-learning models for materials property prediction.\cite{Naik_Dhamrait_Ueltzen_Ertural_Benner_Rignanese_George_2026,nomad_dataset_lobster} The study focused on elastic and phonon properties. In particular, Random Forest and MODNet models incorporating quantum-chemical bonding descriptors were compared with models based entirely on composition and structure-derived features. Beyond improving predictive accuracy, including bonding descriptors enabled a quantitative assessment of the information content of different bonding representations. Feature-importance analyses identified chemically meaningful relationships between quantum-chemical bonding features and target properties, showing that electronic-structure descriptors capture aspects of interatomic interactions not readily encoded in standard geometric feature sets.\cite{Naik_Dhamrait_Ueltzen_Ertural_Benner_Rignanese_George_2026} For properties such as projected force constants and lattice thermal conductivity, these descriptors improved prediction and revealed strong, chemically intuitive correlations between bonding metrics and material behavior.\cite{Naik_Dhamrait_Ueltzen_Ertural_Benner_Rignanese_George_2026} 

More broadly, the study established a general quantitative framework for evaluating bonding descriptors through their predictive utility, a strategy that can readily be extended to alternative bonding representations.

Beyond this, the study demonstrated the value of quantum-chemical bonding descriptors and statistics on them for conventional machine-learning models; their relationship to graph neural networks (GNNs) remains largely unexplored. In particular, it is unclear to what extent the latent representations learned by GNN-based machine learning interatomic potentials (MLIPs) for potential energy surfaces and related materials properties already capture chemically meaningful bonding information. Conversely, it remains an open question whether explicitly incorporating bonding descriptors provides a useful inductive bias that improves learning, generalization, or interpretability. We address both questions through new experiments in Sections \ref{sec:mlips_bondinginfo} and \ref{sec:qc_inclusion_gnn}, respectively.

\section{Do foundation MLIPs implicitly capture bonding information?}
\label{sec:mlips_bondinginfo}

Latent spaces of several foundation MLIPs for materials have previously been shown to converge to a shared chemical space,\cite{Li_Walsh_2026} but whether they also encode information captured by quantum-chemical bonding descriptors remains largely unclear.  While related connections between learned representations and chemical concepts have occasionally been explored in molecular machine-learning models, corresponding studies for foundation MLIPs in crystalline materials remain scarce. To address this central question, we choose two representative foundation models: Allegro and MACE. While both models ultimately express the total energy as a sum of atomic contributions, they differ in their underlying representations. Allegro predicts bond energies before aggregating them into atomic energies, whereas MACE relies on atom-centered energetic representations throughout. By analyzing edge energies and latent representations, we assess whether chemically meaningful bonding information emerges without explicit supervision.

\phantomsection
\addcontentsline{toc}{subsection}{Allegro}
\subsection*{Allegro}

\hfill \break
As we previously mentioned in Sec.~\ref{sec:bonding_amb}, Allegro \cite{musaelian_learning_2023} decomposes the total energy into edge-wise energies ($E_{ij}$), which naturally provides us with bond-level embeddings, unlike other MLIPs that only sum or pool atomic (node-level) energies into the total energy, such as MACE \cite{batatia_mace_2022, Batatia_Batzner_Kovács_Musaelian_Simm_Drautz_Ortner_Kozinsky_Csányi_2022} and NequIP \cite{Batzner_Musaelian_Sun_Geiger_Mailoa_Kornbluth_Molinari_Smidt_Kozinsky_2022}. Hattori et al. \cite{hattori2026chemical} interpreted $E_{ij}$s from Allegro as dissociation energies of their respective bond type, and the authors demonstrated that their trained Allegro model on molecules from the SPICE2 dataset \cite{Eastman_Pritchard_Chodera_Markland_2024, Eastman_Behara_Dotson_Galvelis_Herr_Horton_Mao_Chodera_Pritchard_Wang_etal_2023} implicitly learned the dissociation energy of various bond types from $E_{ij}$ when compared with the literature. Shimamura et al. \cite{Shimamura_Nomura_2026} carried out quantum-chemical calculations on molecules and applied the Interacting Quantum Atoms (IQA) method to partition the total energy into intra- and interatomic contributions. They remarkably found a strong correlation between Allegro $E_{ij}$ and IQA two-body energies, despite not training on them, suggesting that Allegro learns information correlating with meaningful chemical-bonding descriptors. These studies have shown for molecular systems that MLIPs can inherently learn chemical rules and information without direct supervision or explicit molecular input. The following investigation will focus on materials instead.

\begin{figure}[pos=h]
    \centering
    \includegraphics[width=0.68\textwidth]{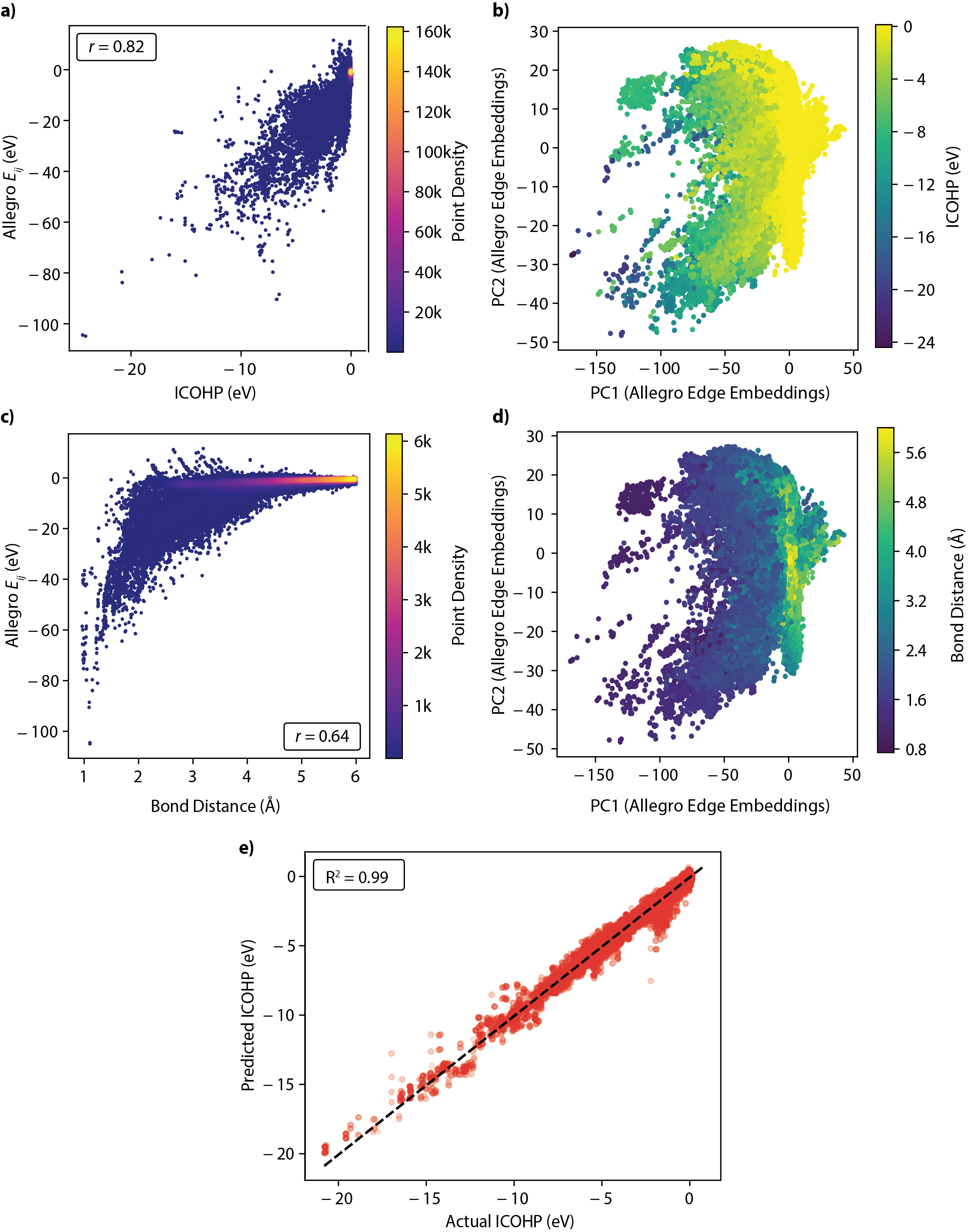}
    \caption{Investigation of Allegro Edge-wise energies (E$_{ij}$) and their embeddings ($x_{ij}$) extracted from the pretrained model "Allegro-OAM-L-0.1" from a bonding perspective. Plot of E$_{ij}$ against a) ICOHP and c) bond distance, with r showing the Pearson correlation between them, color-coded with the density of bond samples. Principal Component Analysis (PCA) of $x_{ij}$ yielded a 2D map of $\mathrm{PC1}$ and $\mathrm{PC2}$ projections of $x_{ij}$, color-coded by b) ICOHP and d) bond distance. e) Parity plot of true vs. predicted  ICOHP using linear probing of Allegro's edge latent representations (i.e., $x_{ij}$). The high correlation indicates that Allegro's latent space encodes and implicitly learns such bonding information. ICOBI results are shown in Fig. \ref{fig:SI_allegro}.}
    \label{fig:Allegro_ICOXX}
\end{figure}

Here, we directly extract $E_{ij}$ from the pretrained Allegro model "Allegro-OAM-L-0.1" \cite{musaelian_learning_2023, Batzner_Musaelian_Sun_Geiger_Mailoa_Kornbluth_Molinari_Smidt_Kozinsky_2022, Tan_Descoteaux_Kotak_De_Miranda_Nascimento_Kavanagh_Zichi_Wang_Saluja_Hu_Smidt_et_al_2026, Musaelian_Johansson_Batzner_Kozinsky_2023} of 5,849,142 edges from 5,978 non-metallic crystalline materials from Materials Project  \cite{Jain_Ong_Hautier_Chen_Richards_Dacek_Cholia_Gunter_Skinner_Ceder_et_al_2013} using a cutoff radius of 6 \si{\angstrom}. We then compare those $E_{ij}$s with ICOHP and ICOBI computed by \cite{Naik_Dhamrait_Ueltzen_Ertural_Benner_Rignanese_George_2026} along with bond distance. Fig.~\ref{fig:Allegro_ICOXX} a) and c) show the correlation of Allegro models $E_{ij}$s with ICOHP and distance, respectively. Comparisons of $E_{ij}$s with ICOBIs are provided in Fig.~\ref{fig:SI_allegro} (a). These panels are color-coded with the data point densities.

Fig.~\ref{fig:Allegro_ICOXX} a) shows that ICOHP decreases with $E_{ij}$, suggesting a strong relationship, as quantified by a high Pearson correlation ($r = 0.82$). We note that ICOHP partitions the bond energy (i.e., the sum of Kohn-Sham orbital energies), not the total energy, as Allegro does.\cite{dronskowski_computational_2005} Thus, we cannot expect a perfect correlation. Additionally, because of the Allegro model's architecture, the pairwise energy contributions are generally not symmetric, i.e., $E_{ij}  \neq E_{ji}$. ICOBI exhibits a similarly strong linear relationship, but with the opposite sign ($r=-0.82$) (Fig.~\ref{fig:SI_allegro} a). The color-coded bond samples cluster near zero, where $E_{ij}$ and ICOHP/ICOBI are zero, consistent with insignificant interactions between $E_{ij}$ and bond energy and bond order. This indicates that Allegro's inferred $E_{ij}$s encode information about quantum-chemical bond strengths and bond orders (i.e., ICOHPs and ICOBIs). 

By contrast, although bond distance is explicitly provided as a geometrical input descriptor for the graph edges, its Pearson correlation ($r = 0.64$) with $E_{ij}$ is lower than that of ICOHP and ICOBI. Moreover, bond samples tend to cluster near $E_{ij}$ $\sim$  0 at larger distances, indicating smaller contributions and weaker interactions. The stronger correlations of $E_{ji}$ with ICOHP and ICOBI than with bond distance suggest that $E_{ij}$ is not merely a learned transformation of geometric information. Instead, through the nonlinear transformations performed during message passing, Allegro appears to learn representations that reflect chemically meaningful bonding information while being trained only on total energies.

After analyzing $E_{ij}$ with respect to ICOHP, ICOBI, and bond distance and identifying the novel correlations, we analyze Allegro's 1,024-dimensional latent representations, specifically the edge embeddings ($x_{ij}$). We directly extract the final-layer edge embeddings ($x^{ij,L=N_{\mathrm{layer}}}$ in the original notation of Ref.\cite{musaelian_learning_2023}), which Allegro subsequently transforms into $E_{ij}$ through a multi-layer perceptron. For simplicity, we denote these embeddings as $x_{ij}$ throughout this work.

Due to the multidimensionality of $x_{ij}$, we perform principal component analysis (PCA) on normalized $x_{ij}$ to visualize the $x_{ij}$ latent space in a two-dimensional ($\mathrm{PC2}$ against $\mathrm{PC1}$) map, with bonds color-coded by ICOHP, ICOBI, and bond distance, as shown in  Fig.~\ref{fig:Allegro_ICOXX} b), Fig.~\ref{fig:SI_allegro} b), and  Fig.~\ref{fig:Allegro_ICOXX} d), respectively. $\mathrm{PC1}$ captures most variation in $x_{ij}$, with an explained variance of 0.07 (7\%) from the 1,024-dimensional $x_{ij}$ embeddings, while $\mathrm{PC2}$ captures the second-largest, at 0.05 (5\%). Although these two components account for only about 12\% of the total variance, they reveal clear structure within the latent space.

From Fig.~\ref{fig:Allegro_ICOXX} b), Fig.~\ref{fig:SI_allegro} b), and  Fig.~\ref{fig:Allegro_ICOXX} d), we see that the 2D map of $x_{ij}$ PCs does not form distinct clusters but exhibits a gradual change, accompanied by a continuous shift in color representing ICOHP, ICOBI, and bond distance. This indicates a correlation among ICOHP, ICOBI, and bond distance, with clear variation in $x_{ij}$ $\mathrm{PC1}-\mathrm{PC2}$. Even though visually such a gradient change is also observed in the bond distance plot in Fig.~\ref{fig:Allegro_ICOXX} d), the maximum bond distance in yellow is not as dominant in the far right of the 2D map as seen in the maximum values of ICOHP and ICOBI, suggesting better alignment of the $x_{ij}$ $\mathrm{PC1}$ projection with ICOHP and ICOBI compared to bond distance. This is also shown in $\mathrm{dCor}$ between $x_{ij}$ $\mathrm{PC1}-\mathrm{PC2}$ and the other bonding descriptors in Table~\ref{tab:Allegro_Eij_cor}, where ICOHP and ICOBI show higher correlation of 0.61 and 0.59, respectively, compared to a dCor of 0.46 between $x_{ij}$ $\mathrm{PC1}-\mathrm{PC2}$ and bond distance. 

The lower $\mathrm{dCor}$ values of Allegro's $x_{ij}$ $\mathrm{PC1}-\mathrm{PC2}$ with the other bonding descriptors than those obtained directly for  $E_{ij}$ (Table~\ref{tab:Allegro_Eij_cor}) are expected because only ~12\% of the variance in $x_{ij}$ is captured by the first two PC projections, leading to information loss during dimensionality reduction. Nevertheless, although $\mathrm{PC1}$ captures only about 7\% of the variance, the data suggest that as $\mathrm{PC1}$ decreases, ICOHP also decreases, while ICOBI increases. Overall, this shows that the direction of greatest variance in $x_{ij}$ is consistent with ICOHP and ICOBI because lower $E_{ij}$ is accompanied by lower ICOHP and higher ICOBI. Taken together, this indicates that Allegro's latent representations encode bonding information without direct supervision or explicit instructions, simply by optimizing its weights to predict electronic-structure properties, namely energy and forces. 

\begin{table}[pos=h]
\centering
\caption{Distance correlation ($\mathrm{dCor}$) coefficients between bonding descriptors and Allegro outputs of $E_{ij}$ and $\mathrm{PC1}-\mathrm{PC2}$($x_{ij}$). Pearson correlation ($r$) is computed only between the bonding descriptors and $E_{ij}$ because it can be computed only between two single-dimensional vectors, which explains the ------ in their respective space.}
\begin{tabular}{lcccc}
\toprule
\multirow{2}{*}{\textbf{Bonding Descriptor}} &
\multicolumn{2}{c}{\textbf{Pearson Correlation ($r$)}} &
\multicolumn{2}{c}{\textbf{Distance Correlation ($\mathrm{dCor}$)}} \\
\cmidrule(lr){2-3} \cmidrule(lr){4-5}
&
$E_{ij}$ &
$\mathrm{PC1}-\mathrm{PC2}\!\left(x_{ij}\right)$ &
$E_{ij}$ &
$\mathrm{PC1}-\mathrm{PC2}\!\left(x_{ij}\right)$ \\
\midrule
ICOHP (eV)          & 0.82  & ------ & 0.82 & 0.61 \\
ICOBI               & -0.82 & ------ & 0.81 & 0.59 \\
Bond distance (\AA) & 0.64  & ------ & 0.68 & 0.46 \\
\bottomrule
\end{tabular}
\label{tab:Allegro_Eij_cor}
\end{table}

To further assess whether chemical bonding information is encoded in MLIP latent representations, we perform linear probing to quantify how much information about ICOHP and ICOBI is already present or implicitly learned in the MLIP latent space. To observe this effect, we fit a linear regression model on the raw $x_{ij}$ (without normalization) to predict ICOHP and ICOBI, using 80\% of the dataset for training and 20\% for testing, with seed 42 for splitting. To prevent data leakage from splitting by edges in the train-test split, we group materials by their reduced formula, then perform the split. This prevents edges from the same local environment from appearing in both the train and test sets. However, splitting by material structures also does not cause data leakage, as our dataset contains no duplicate or near-duplicate material structures. We show the test set results in Fig.~\ref{fig:Allegro_ICOXX} e) and Fig.~\ref{fig:SI_allegro} c). High coefficients of determination ($R^2$) of 0.99 and 0.98 for ICOHP and ICOBI, respectively, show that both bond energy and bond order are linearly accessible from Allegro's latent representations. This further indicates that Allegro's latent representations implicitly learn bonding-related properties during training.


\phantomsection
\addcontentsline{toc}{subsection}{MACE}
\subsection*{MACE}


Having shown that Allegro’s edge latent representations implicitly learn ICOHP and ICOBI, we next examine whether this also holds for other foundation MLIPs with a different architecture such as MACE. For this, we analyze the node embeddings of MACE, extracted from the pretrained MACE-MPA-0 (medium) model\cite{Batatia_Batzner_Kovács_Musaelian_Simm_Drautz_Ortner_Kozinsky_Csányi_2022, batatia_mace_2022, batatia_foundation_2025, Yuan_Chiang, riebesell2025framework} for the same 5,978 materials used in the Allegro analysis (108,931 nodes in total). Because MACE embeddings before the readout are node-wise, we generate corresponding site-wise (node-wise) features from LOBSTER data, including atomic charges and ICOHP and ICOBI statistics. We compute the $\mathrm{dCor}$ between these site-wise LOBSTER descriptors and MACE node embeddings and find a high $\mathrm{dCor}$ of 0.78.  This high correlation suggests that MACE latent representations encode local chemical-environment information such as charge, bond energy, and bond order. 

\begin{figure}[pos=h]
    \centering
    \includegraphics[width=0.65\linewidth]{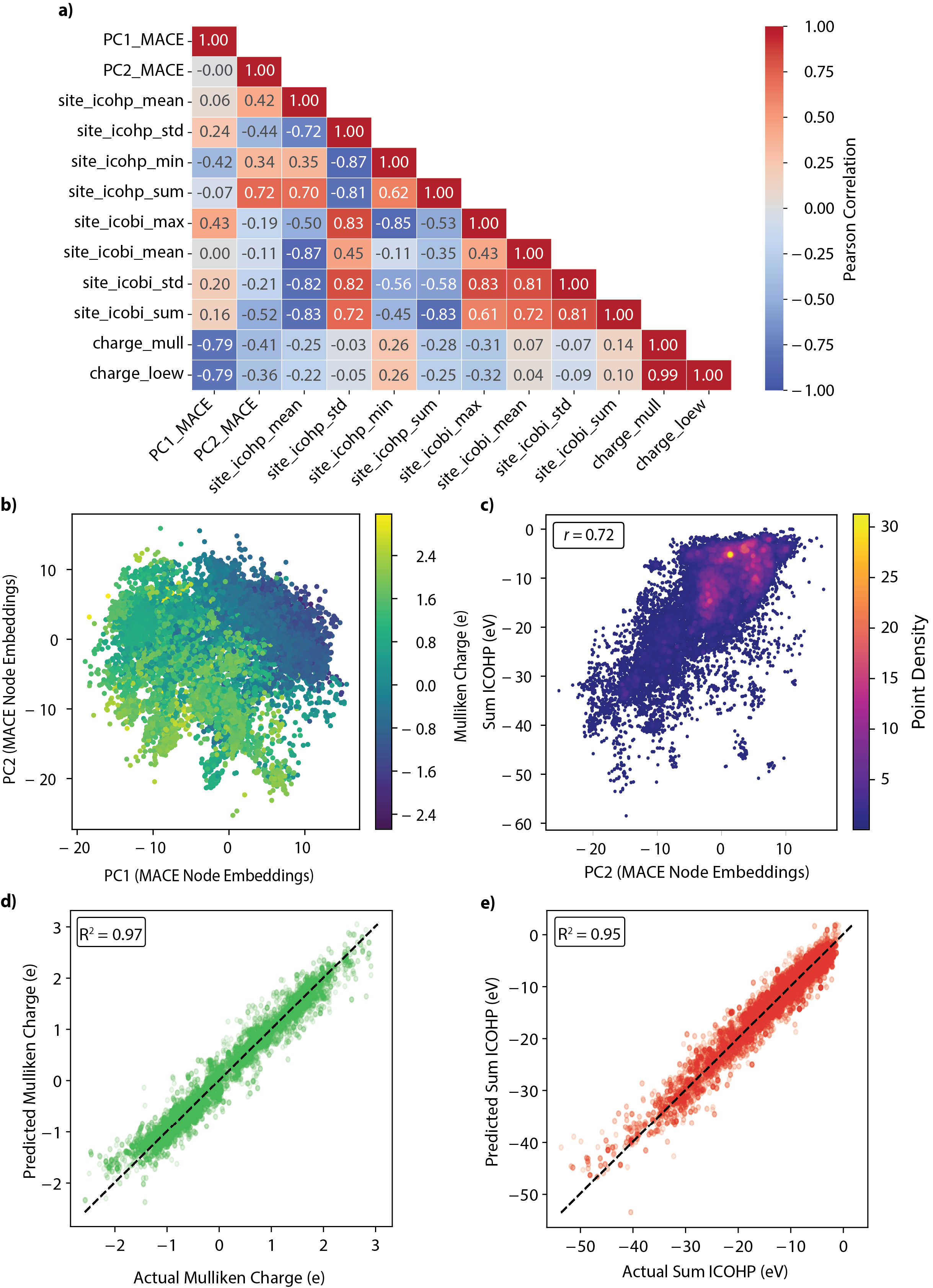}
    \caption{Investigation of the correlations between MACE latent representations and quantum-chemical bonding descriptors. a) Pairwise correlations through a Pearson correlation matrix between the first two projected principal components ($\mathrm{PC1}$ and $\mathrm{PC2}$) of the MACE node embeddings and the generated bonding descriptors. b) 2D map of PC projections on the MACE node embeddings color-coded by Mulliken charge, exhibiting a charge gradient aligned with $\mathrm{PC1}$. c) Correlation between $\mathrm{PC2}$ and summed ICOHP ($r = 0.72$) and is color-coded with the density of site samples. Corresponding Löwdin results are shown in Fig. ~\ref{fig:SI_mace_loew_results}, and correlations between $\mathrm{PC1}$ and Mulliken and between $\mathrm{PC2}$ and ICOBI are shown in Fig. \ref{fig:SI_mace}. So far, these results indicate that $\mathrm{PC1}$ primarily encodes charge information, whereas $\mathrm{PC2}$ encodes bond-energy information about local chemical environments.  d) and e) Parity plots of true versus predicted Mulliken charge and summed ICOHP, respectively, are obtained by linear probing of the MACE latent representations. The high correlation indicates that the MACE latent space encodes and implicitly learns charge and bonding information of the local chemical environment. ICOBI results are shown in Fig. \ref{fig:SI_mace}.}
    \label{fig:DRCor_MACE}
\end{figure}

We then project the normalized MACE 256-dimensional node embeddings onto the first 2 PCs (i.e., $\mathrm{PC1}-\mathrm{PC2}$ only) to compute pairwise Pearson correlations between MACE $\mathrm{PC1}$ and MACE $\mathrm{PC2}$ separately with our generated bonding descriptors. We will use these bonding descriptors later to visualize the descriptors most strongly correlated with those PCs. The projected MACE $\mathrm{PC1}-\mathrm{PC2}$ capture explained variances of 0.24 (24\%) and 0.13 (13\%), respectively, accounting for 37\% of the total variance in the MACE latent representations. Fig.~\ref{fig:DRCor_MACE} a) shows the pairwise Pearson correlations between $\mathrm{PC1}-\mathrm{PC2}$ and the bonding descriptors.

Among all the descriptors, Mulliken and Löwdin charges exhibit the strongest correlations with $\mathrm{PC1}$, reaching $r = -0.79$. That strong correlation between MACE $\mathrm{PC1}$ and Mulliken charge is visually confirmed by a negative quasi-linear correlation in Fig.~\ref{fig:SI_mace} a). We also observe a weaker correlation between Mulliken charge and MACE  $\mathrm{PC2}$ ($r = -0.41$), indicating that MACE PC2 captures some charge information as well, and together MACE $\mathrm{PC1}-\mathrm{PC2}$ show higher $\mathrm{dCor}$ scores with Löwdin and Mulliken charges of 0.85 and 0.83, respectively, compared to the sole $\mathrm{PC1}$ or $\mathrm{PC2}$ $r$ scores as shown in Table~\ref{tab:charge_corr}.

\begin{table}[pos=h] 
\centering 
\caption{Correlation between atomic charge descriptors and $\mathrm{PC1}$ and $\mathrm{PC2}$ of MACE node embeddings ($h_{i}$). $r$ is computed only between $\mathrm{PC1}$ and $\mathrm{PC2}$ because it can only be computed between two single-dimensional arrays, which explains the separate results for each PC. $\mathrm{dCor}$ can find the correlation between the charge descriptors and the 2D $\mathrm{PC1}-\mathrm{PC2}$ array.} 
\begin{tabular}{lccc} 
\toprule 
\textbf{Atomic Charge Descriptor} 
& \multicolumn{2}{c}{\textbf{Pearson Correlation ($r$)} } 
& \textbf{Distance Correlation ($\mathrm{dCor}$)} \\ 
\cmidrule(lr){2-3} \cmidrule(lr){4-4} 
& $\mathrm{PC1}$($h_{i}$) & $\mathrm{PC2}$($h_{i}$) & $\mathrm{PC1}-\mathrm{PC2}$ ($h_{i}$) \\ 
\midrule 
Löwdin & -0.79 & -0.36 & 0.83 \\ 
Mulliken & -0.79 & -0.41 & 0.85 \\ 
\bottomrule 
\end{tabular} 
\label{tab:charge_corr}
\end{table}

This high $\mathrm{dCor}$ prompts us to examine the relationship of Löwdin and Mulliken charges visually in a 2D map of MACE PCs. As can be seen from Fig.~\ref{fig:DRCor_MACE} b) and Fig.~\ref{fig:SI_mace_loew_results} a), negatively charged atoms tend to occupy the positive $\mathrm{PC1}$ and $\mathrm{PC2}$ region of the MACE node embedding space, whereas the positively charged sites are predominantly located in the opposite region (confirming the negative correlations). Rather than forming sharp and separated clusters in the MACE embedding space, the transition from the positively charged region to the negatively charged one is smooth. This behavior suggests that MACE implicitly learned and encoded underlying chemical information about charge and ionicity in various local chemical environments. Since larger charge magnitudes may indicate greater charge separation and a higher ionic contribution to bonding, the MACE embedding space seems to capture such chemical bonding information. These trends appear despite optimizing MACE model weights for accurate predictions of energy, forces, and stress, without explicitly providing charge, oxidation state, electronegativity, or similar embeddings in the input graphs.

After showing how MACE $\mathrm{PC1}$ captured charge-related information, we now analyze the correlations between MACE $\mathrm{PC2}$ and other bonding descriptors related to ICOHP and ICOBI shown in Fig.~\ref{fig:DRCor_MACE} a). We observe a strong positive correlation between MACE $\mathrm{PC2}$ and the summed ICOHP per site ($r = 0.72$, Fig.~\ref{fig:DRCor_MACE} c) and a somewhat weaker negative correlation with the summed ICOBI per site ($r = -0.52$, Fig.~\ref{fig:SI_mace} b), denoted by site\_icohp\_sum and site\_icobi\_sum, respectively. In contrast, MACE $\mathrm{PC1}$ exhibits little correlation with either descriptor despite accounting for the largest variance in the MACE latent representations.

Together with the strong correlation between $\mathrm{PC1}$ and charge, these results on $\mathrm{PC2}$ and ICOHP/ICOBI suggest that $\mathrm{PC1}$ and $\mathrm{PC2}$ encode different aspects of the local chemical environment. Specifically, $\mathrm{PC1}$ primarily captures charge-related information, whereas $\mathrm{PC2}$ is associated with variations in bonding interactions as quantified by ICOHP and ICOBI. Overall, these findings indicate that the multidimensional MACE latent space implicitly learns and encodes chemically meaningful information about local atomic environments, as reflected by its correlations with charge, ICOHP, and ICOBI.

As in our Allegro latent space analysis, we use the MACE latent representations of the same materials to assess how much charge, ICOHP, and ICOBI information the MACE foundation model learned during training, following \cite{Kolganov_Klein_Pidko_2026}, which recovered chemical information for molecules such as Brønsted and Lewis acidity and ring aromaticity, reporting Spearman correlations of 0.8-0.99. To this end, we linear-probe the raw MACE latent representations without normalization to predict Mulliken charges and the sum of ICOHP and ICOBI by fitting a linear regression model. We follow the same 80\% train/20\% test split with seed 42 for splitting as in our Allegro analysis, fit the model, and evaluate predictions on the test set. As in our Allegro data split, to prevent data leakage from splitting by nodes, we group materials by their reduced formula, then perform the split. This prevents equivalent atoms from the same material from appearing in both the train and test sets. We report the test-set results in Fig.~\ref{fig:DRCor_MACE} d), e), and Fig.~\ref{fig:SI_mace} c). We observe high coefficients of determination (R$^2$) of 0.97, 0.95, and 0.95 via linear probing for Mulliken charge, sum ICOHP, and sum ICOBI, respectively. These results show that bonding properties such as charge, sum of bond energies, and bond order are linearly accessible from the MACE latent space, further confirming that MACE implicitly learns the local chemical environment during training.

\phantomsection
\addcontentsline{toc}{subsection}{Chemical Information in Learned Latent Spaces of MLIPs}
\subsection*{Chemical Information in Learned Latent Spaces of MLIPs}

Taken together, the analyses of Allegro and MACE suggest that modern foundation MLIPs implicitly learn representations that are strongly correlated with established quantum-chemical bonding descriptors. Bond energies, bond orders, and atomic charges are all found to be linearly accessible from the latent spaces of pretrained models, despite never being provided during training. These results suggest that chemically meaningful information emerges naturally when learning accurate potential-energy surfaces. This is perhaps not surprising, as contemporary graph-based MLIPs represent materials through local atomic environments and interatomic interactions, thereby introducing an implicit bonding picture at the architectural level. However, the learned representations remain high-dimensional and are generally not directly interpretable. While they encode bonding-related information, they do not provide the low-complexity, human-readable descriptors needed for chemical reasoning in the same way as bond order, bond strength, or oxidation state. Instead, the results suggest that traditional bonding descriptors and learned latent representations capture overlapping information at different levels of abstraction: the former emphasize interpretability and reasoning, while the latter optimize for prediction. The imperfect correlations observed here further indicate that both representations contain shared as well as complementary information about the underlying electronic structure.

\section{Integration of quantum-chemical bonding descriptors in deep-learning architectures}\label{sec:qc_inclusion_gnn}

The previous section showed that modern foundation MLIPs implicitly learn representations strongly correlated with established quantum-chemical bonding descriptors. Together with earlier observations that such bonding descriptors improve feature-based machine-learning models based solely on structural and compositional features,\cite{Naik_Dhamrait_Ueltzen_Ertural_Benner_Rignanese_George_2026} this raises the question of whether they can also improve graph-based deep-learning models when incorporated explicitly into their representations.

We therefore examine how quantum-chemical bonding information can be integrated into graph neural networks for crystalline materials. A similar strategy has recently been proposed and successfully applied to catalytic-property prediction.\cite{yin_local_2026}

We explicitly integrate quantum-chemical bonding descriptors into a new GNN architecture based on MEGNet \cite{chen_graph_2019}, which we name the electronic-structure-featurized MEGNet (ESF-MEGNet). Although the current implementation focuses on bonding-related descriptors, a broader name reflects the model's possible generalization to arbitrary electronic-structure information. In principle, similar modifications could be implemented in other graph neural network architectures; however, we chose MEGNet as a well-established, conceptually simple testbed to investigate whether explicit bonding information improves graph-based materials models.

\begin{figure}[pos=h]
    \centering
    \includegraphics[width=0.8\linewidth]{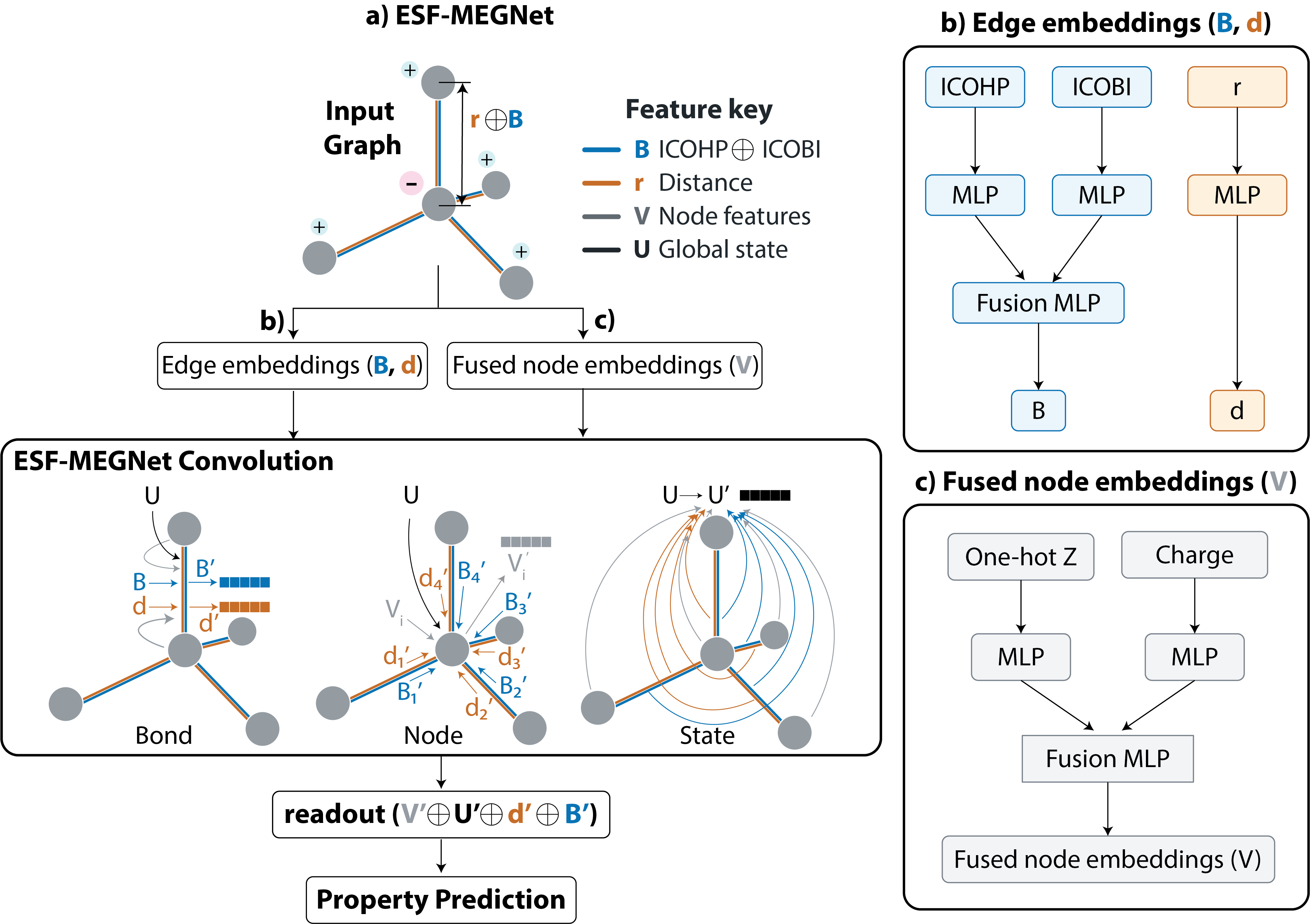}
    \caption{Model architecture of the ESF-MEGNet model. a) shows the overall model architecture, where the graph shows the added electronic-structure features: charge as a node feature, ICOHP and ICOBI as edge features. Then, the edge ($B$ and $d$), node ($V$), and state ($U$) embeddings propagate into the ESF-MEGNet convolution and are updated to $B'$, $d'$, $V'$, and $U'$. Those updated embeddings are aggregated by set2set pooling to later predict material properties. b) the two-channel edge embeddings $B$ and $d$ before passing them to the ESF-MEGNet convolution. c) detailed architecture of fusing the MLP-transformed one-hot encoded atomic number (Z) and MLP-transformed charge features before passing them to the ESF-MEGNet convolution. }
    \label{fig:mod_arch}
\end{figure}

A detailed overview of the ESF-MEGNet architecture is provided in Fig. ~\ref{fig:mod_arch}. As shown in Fig.~\ref{fig:mod_arch} b), node features derived from quantum-chemical bonding analysis (specifically Löwdin charges) are processed separately from the one-hot-encoded atomic number ($Z$) and then combined into the fused node embeddings ($V$) before feeding them into the ESF-MEGNet convolution. These embeddings therefore carry information about both the chemical species and atomic charges from quantum-chemical bonding analysis into the convolution.

For the edge embeddings ($B$ and $d$), we distinguish between geometric information and quantum-chemical bonding information for each bond. As shown in Fig.~\ref{fig:mod_arch}c, the Gaussian-expanded bond-length vector ($r$) is transformed using a multi-layer perceptron ($\mathrm{MLP}$), separately from ICOHP and ICOBI, which are passed through individual MLPs before being fused into $B$ (i.e., $B = \mathrm{MLP}(\mathrm{ICOHP}) \oplus \mathrm{MLP}(\mathrm{ICOBI})$).

This separation, or two-channel design, for geometric and quantum-chemical bonding information is motivated by the different information ICOHP and ICOBI encode, commonly interpreted as measures of bond strength and bond order, respectively. During message passing in the MEGNet convolution block in Fig.~\ref{fig:mod_arch} a), edge features from ICOHP and ICOBI are updated separately from the bond-length features but with equivalent operations. Specifically, the new edge embeddings (either $B'$ or $d'$) are updated by concatenating the original edge embeddings ($B$ or $d$), state embeddings ($U$), and node embeddings from the source ($V_s$) and destination nodes ($V_r$), then transforming the concatenated embeddings with an MLP, as shown in equations \ref{eq:separate_edge_updates} below:

\begin{equation} \begin{aligned} \mathbf{d}'_{k} &= \phi_{d}\left( \mathbf{v}_{s_k} \oplus \mathbf{v}_{r_k} \oplus \mathbf{d}_{k} \oplus \mathbf{u} \right), \\[4pt] \mathbf{B}'_{k} &= \phi_{B}\left( \mathbf{v}_{s_k} \oplus \mathbf{v}_{r_k} \oplus \mathbf{B}_{k} \oplus \mathbf{u} \right), \end{aligned} \label{eq:separate_edge_updates} \end{equation}
where $\phi$ is the activation function. 

The resulting edge embeddings $d'$ and $B'$ are subsequently incorporated into the node ($V$) and state ($U$) updates. For the node update, the original $V$ and $U$ embeddings, together with the updated edge embeddings $B'$ and $d'$, are concatenated and transformed by an MLP to obtain the updated node embeddings $V'$, as shown in Eq.~\ref{eq:node_update}:

\begin{equation}
\mathbf{v}_i' = \phi\left(\mathbf{v}_i^{d'} \oplus \mathbf{v}_i^{B'} \oplus \mathbf{v}_i \oplus \mathbf{u}\right)
\label{eq:node_update}
\end{equation}

Similarly, the updated edge embeddings $B'$ and $d'$, together with the updated node embeddings $V'$, are incorporated into the state update. Specifically, the original state embeddings $U$ are concatenated with $V'$, $B'$, and $d'$ and subsequently transformed by an MLP to obtain the updated state embeddings $U'$, as shown in Eq.~\ref{eq:state_update}:

\begin{equation}
\mathbf{u}' = \phi\left(\bar{\mathbf{u}}^{d'} \oplus \mathbf{u}^{B'} \oplus \bar{\mathbf{u}}^{v'} \oplus \mathbf{u}\right)
\label{eq:state_update}
\end{equation}

In the ESF-MEGNet convolution, each of the four feature types ($d$, $B$, $V$, and $U$) is updated using information from the others, except for the edge representations $d$ and $B$, which are updated only from the node ($V$) and state ($U$) embeddings. The updated representations $d'$, $B'$, $V'$, and $U'$ are subsequently aggregated using set2set pooling to predict material properties. While the specific architectural modifications introduced here are not unique and could be implemented in different ways, our primary objective is to investigate whether incorporating quantum-chemical bonding descriptors in structure-based graph representations improves material-property prediction.

To assess the significance of the added bonding features, we compare the performance of ESF-MEGNet with that of the baseline MEGNet model for several target properties: the last phonon DOS peak frequency,  mean-squared displacement averaged over non-equivalent sites (Avg. MSD), bulk modulus ($K$), shear modulus ($G$), and Vickers hardness ($H_V$). For some of them, we have previously seen improved performance when comparing feature-based models to those based solely on structural and compositional features.\cite{Naik_Dhamrait_Ueltzen_Ertural_Benner_Rignanese_George_2026}

\begin{table}[pos=hb] 
\centering 
\caption{Average test set performance metrics of MEGNet and ESF-MEGNet variants on different target properties across all 10 cross-validation k-folds. Results are reported as the mean ($\mu$) and standard deviation ($\sigma$) calculated from the fold-wise test metrics. The best-performing model, ESF-MEGNet, is shown in bold across all material properties. The MAE improvement (\%) shown beside ESF-MEGNet MAEs is calculated with respect to the baseline MEGNet model as shown in equation \ref{eq:mae_improvement}.} \label{tab:model_comparison}  
\begin{tabular}{llcc} \toprule \textbf{Property} & \textbf{Model} & \textbf{Avg. MAE} & \textbf{Avg. $R^2$} \\ \midrule 
\multirow{4}{*}{Last phonon DOS peak (cm$^{-1}$)} 
&ESF-MEGNet & \textbf{33.728 $\pm$ 4.795 (9\%)} & 0.982 $\pm$ 0.008 \\ 
&ESF-MEGNet-rand & 35.518 $\pm$ 6.832 & \textbf{0.983 $\pm$ 0.004} \\ 
&ESF-MEGNet-zero & 37.049 $\pm$ 7.917 & 0.980 $\pm$ 0.008 \\ 
&MEGNet & 36.926 $\pm$ 7.967 & 0.980 $\pm$ 0.009 \\
\midrule 

\multirow{4}{*}{Avg. MSD, $\log_{10}(\text{\AA}^2)$} 
&ESF-MEGNet & \textbf{0.075 $\pm$ 0.005  (4\%)} & \textbf{0.879 $\pm$ 0.029} \\ 
&ESF-MEGNet-rand & 0.080 $\pm$ 0.005 & 0.855 $\pm$ 0.026 \\ 
&ESF-MEGNet-zero & 0.081 $\pm$ 0.007 & 0.848 $\pm$ 0.035 \\ 
&MEGNet & 0.078 $\pm$ 0.007 & 0.867 $\pm$ 0.034 \\
\midrule 

\multirow{4}{*}{Bulk modulus $K$, $\log_{10}(\mathrm{GPa})$} 
&ESF-MEGNet & \textbf{0.081 $\pm$ 0.007 (15\%)} & \textbf{0.867 $\pm$ 0.069} \\ 
&ESF-MEGNet-rand & 0.098 $\pm$ 0.011 & 0.843 $\pm$ 0.073 \\ 
&ESF-MEGNet-zero & 0.097 $\pm$ 0.010 & 0.832 $\pm$ 0.078 \\ 
&MEGNet & 0.095 $\pm$ 0.009 & 0.833 $\pm$ 0.065 \\
\midrule 

\multirow{4}{*}{Shear modulus $G$, $\log_{10}(\mathrm{GPa})$} 
&ESF-MEGNet & \textbf{0.095 $\pm$ 0.006  (15\%)} & \textbf{0.859 $\pm$ 0.060} \\ 
&ESF-MEGNet-rand & 0.120 $\pm$ 0.014 & 0.813 $\pm$ 0.083 \\ 
&ESF-MEGNet-zero & 0.118 $\pm$ 0.010 & 0.802 $\pm$ 0.063 \\ 
&MEGNet & 0.112 $\pm$ 0.010 & 0.827 $\pm$ 0.066 \\
\midrule 

\multirow{4}{*}{Vickers hardness $H_V$, $\log_{10}(\mathrm{GPa})$} 
&ESF-MEGNet & \textbf{0.122 $\pm$ 0.007 (12\%)} & \textbf{0.765 $\pm$ 0.044} \\ 
&ESF-MEGNet-rand & 0.142 $\pm$ 0.007 & 0.703 $\pm$ 0.045 \\ 
&ESF-MEGNet-zero & 0.137 $\pm$ 0.005 & 0.727 $\pm$ 0.030 \\ 
&MEGNet & 0.139 $\pm$ 0.009 & 0.708 $\pm$ 0.044 \\
\bottomrule 
\end{tabular} 
\end{table}

To disentangle the effects of the bonding descriptors from the modified architecture, we also consider two control models derived from ESF-MEGNet. In ESF-MEGNet-zero, the charge, ICOHP, and ICOBI features are replaced by zeros, whereas in ESF-MEGNet-rand they are replaced by random values. Because both control models retain the same architecture as ESF-MEGNet, any performance differences can be attributed to the information in the bonding descriptors rather than to architectural changes or increased model complexity alone. 

We employ 10-fold cross-validation using an 80\%/10\%/10\% train/validation/test split in each fold. Model performance is evaluated using the mean absolute error (MAE) and $R^2$. Table~\ref{tab:model_comparison} reports the average MAE and $R^2$ across all folds, while Tables~\ref{tab:cv_results_last_peak},~\ref{tab:cv_results_avgMSD},~\ref{tab:cv_results_K},~\ref{tab:cv_results_G},~\ref{tab:cv_results_Hv} in the Supporting Information provide detailed fold-by-fold results for individual folds.

As shown by the MAE improvement in percent in ESF-MEGNet rows of Table~\ref{tab:model_comparison}, ESF-MEGNet achieves lower MAEs (and higher $R^2$ values) across all target properties compared with both the baseline MEGNet model and the corresponding control models ESF-MEGNet-zero and ESF-MEGNet-rand. Because the control models retain the same architecture as ESF-MEGNet but remove the chemical information in the charge, ICOHP, and ICOBI descriptors, these results indicate that the observed performance gains arise primarily from the additional bonding information rather than architectural modifications alone. Explicitly incorporating quantum-chemical bonding descriptors therefore provides information that is not fully captured by structure-based graph representations. These findings across diverse material properties suggest that similar featurization strategies may also prove beneficial in other graph-based models for material property prediction.

\begin{figure}[pos=ht]
    \centering
    \includegraphics[width=0.7\textwidth]{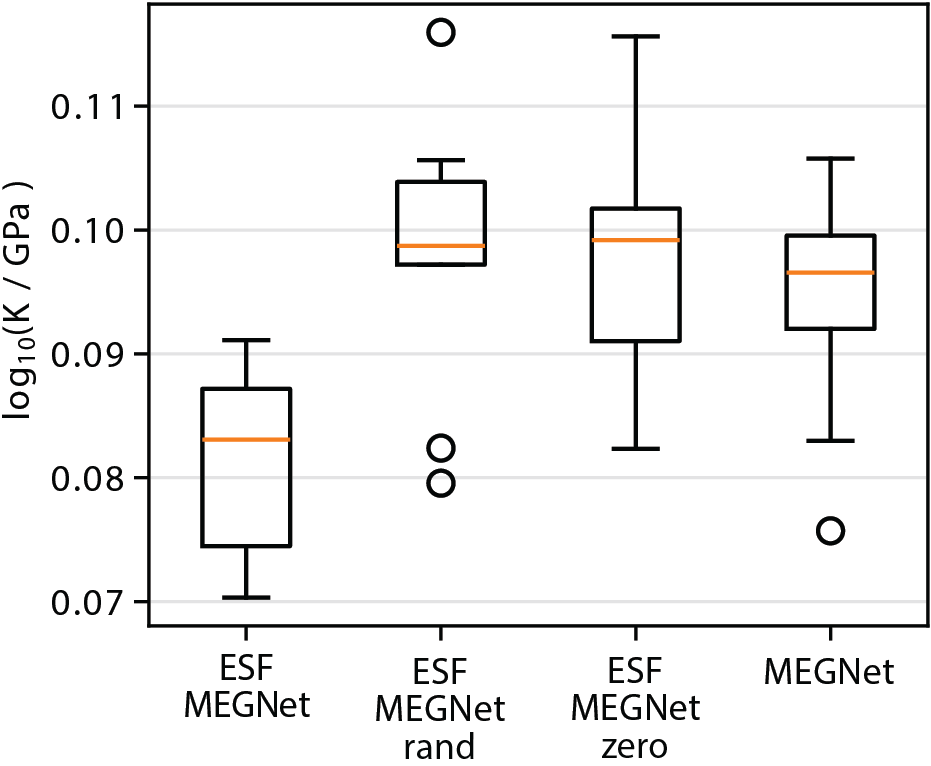}
    \caption{These box plots show the distribution of the MAEs across all ten folds from the four models predicting K. The MAE distribution shows that ESF-MEGNet outperforms the baseline MEGNet and the two control models, ESF-MEGNet-zero and ESF-MEGNet-rand, confirming that the quantum-chemical bonding descriptors add information that improves material property predictions.
    }
    \label{fig:K_boxPlots}
\end{figure}

Visually, the box plots in Fig.~\ref{fig:K_boxPlots} and Fig.~\ref{fig:SI_boxPlots} show that ESF-MEGNet achieves a significantly lower MAE than the baseline MEGNet and other control models for several properties, including bulk and shear moduli, hardness, and Avg. MSD. The MAE improvement is smaller for the last phonon DOS peak. This may be because some of these properties have been shown to correlate with the chemical bonding descriptors used in our model. For example, bond energy and ionicity have been reported to empirically predict the bulk modulus of various materials.\cite{Xu_Wang_Tian_2013, Muttalib_Abdullah_2025} Shear modulus is correlated with Valence electron concentration.\cite{Jhi_Ihm_Louie_Cohen_1999} Morevoer, It has been reported that covalent bond number,\cite{Phillips_1970} bond strength,\cite{Simunek_Vackar_2006} and bond overlap population\cite{Guo_Li_Liu_Yu_He_Liu_Xu_Tian_Wang_2008} are correlated with hardness. Furthermore, Al-Fahdi et al. \cite{Al_Fahdi_Rurali_Hu_Wolverton_Hu_2025} showed that MSD correlates with ICOHP and ICOBI. Moreover, Cs in \ce{Cs3Bi2I9} shows the largest MSD and the weakest bonds, as evidenced by low |COHP| values between Cs and neighboring sites,\cite{Acharyya_Pal_Ahad_Sarkar_Rana_Dutta_Soni_Waghmare_Biswas_2023} consistent with other studies linking thermal properties to bonding in Refs. \cite{He_Xia_Lin_Pal_Zhu_Kanatzidis_Wolverton_2022, Chang_Ma_Yuan_Zheng_Wei_Al-Fahdi_Gao_Zhang_Shao_Hu_etal._2022, Al-Fahdi_Zhang_Hu_2021, Yuan_Chen_Liao_2023, Wuttig_Deringer_Gonze_Bichara_Raty_2018}. Therefore, adding ICOHP, ICOBI, and charge descriptors to the ESF-MEGNet model likely played a major role in significantly improving these material property predictions. 

We also note that the inclusion of explicit angular representations in models such as CHGNet,\cite{Deng_Zhong_Jun_Riebesell_Han_Bartel_Ceder_2023} ALIGNN,\cite{Choudhary_DeCost_2021} M3GNet,\cite{chen_universal_2022} and MatRIS\cite{Zhou_Hu_Zhang_Wang_Tan_Jia_2026} would not preclude the incorporation of quantum-chemical bonding descriptors, since these architectures still retain conventional atom-centered message-passing pathways. The present work demonstrates that ICOHP- and charge-based descriptors provide information complementary to that contained in interatomic distances alone. Although message-passing models built from pairwise geometric inputs can represent many-body dependencies, their expressivity and data efficiency depend on the architecture. Whether the benefits of the added bonding descriptors persist in models with explicit angular or higher-order correlations remains to be tested. Answering this question would require disentangling the effects of the bonding descriptors from those of the underlying model architecture and therefore lies beyond the scope of the present work. 

Taken together, these results show that quantum-chemical bonding descriptors can complement structural graph representations and improve predictions of a range of materials properties. Combined with the observation that modern foundation MLIPs already learn representations strongly correlated with bonding descriptors, the findings suggest substantial overlap, but not complete redundancy, between learned and chemically motivated representations. More specifically, the present work demonstrates that bonding descriptors contribute information beyond that contained in pairwise interatomic distances. Whether similar gains can be achieved in architectures that explicitly encode many-body interactions remains an open question. This interplay of shared and complementary information motivates the future directions discussed in the following sections.

\section{Future directions}

Having established the relationship between bonding descriptors and deep-learning MLIPs, we now explore future directions that build on their correlation and complementarity. We focus on directions that we consider particularly promising, while emphasizing that the potential applications of bonding descriptors are unlikely to be limited to the examples discussed here.

\subsection{Interpretable Models for Property Prediction Based on Foundational Bonding Models}

The studies discussed above show that quantum-chemical bonding information complements structural and compositional representations and can improve material-property predictions. Thus, they can, in part, carry more information than typical compositional and structural features, making them especially attractive for interpretable machine learning models. Further, bonding descriptors provide links to established chemical concepts and enable machine-learning predictions to be rationalized. Interpretable models are highly desirable as they "increase trust in black-box methods, help reduce errors, allow for the improvement of the underlying models, enhance human-AI collaboration, and ultimately enable fully automated scientific discoveries that remain understandable to human scientists."\cite{wetzel_interpretable_2025} Thus, bonding descriptors offer a promising route toward machine-learning models that are both predictive and chemically interpretable. We review the role of bonding descriptors in this context and provide ideas for future directions.

The need to perform electronic-structure calculations and subsequent bonding analyses limits the scalability and widespread use of such descriptors for chemical interpretation and rationalization. Encouragingly, previous work has shown that bonding descriptors can be predicted accurately from alternative materials representations.\cite{Naik_Dhamrait_Ueltzen_Ertural_Benner_Rignanese_George_2026} In this work, we show that latent representations learned by machine-learning interatomic potentials (MLIPs) can also predict quantum-chemical bonding descriptors. Taken together, these findings suggest that bonding information can be encoded in such materials representations and can be recovered without performing explicit bonding analyses. Given sufficiently large datasets, this opens the possibility of developing foundation models for quantum-chemical bonding descriptors, with potential transferability across different levels of electronic-structure theory.

Bonding data, either from large-scale databases or a bonding foundation model, can be used to extract chemically meaningful relationships underlying material properties.
We present a proof of concept from our previous work \cite{Naik_Dhamrait_Ueltzen_Ertural_Benner_Rignanese_George_2026}, which we summarize here. Combining our chemical bonding database with symbolic regression, a methodology that identifies explicit analytical expressions directly from data, enabled us to derive simple, chemically intuitive expressions that explain the variance well. 

Using an All Relevant Feature Selection method (ARFS, GrootCV in our case), we selected the 20 most relevant features for subsequent symbolic regression. The Sure Independence Screening Sparsifying Operation (SISSO)\cite{ouyang_sisso_2018,Purcell2022,purcell2023recent} algorithm then found the best n-dimensional (n=1,2) descriptor from these 20 features using predefined mathematical operators. We applied this methodology to a range of elastic and phonon-related properties, for which predictive performance improved substantially upon inclusion of bonding information in the descriptor space. Below, we highlight representative examples that illustrate both the predictive value and chemical interpretability of the resulting descriptors.

The first example concerns the maximal projected force constant per material ("$\mathrm{max\_pfc}$"), which reflects the stiffness of the strongest interatomic interactions and influences a wide range of vibrational properties. The force constants were obtained from harmonic phonon calculations and projected onto individual bonds. Using the methodology above on a diverse dataset of 1,264 projected force constants, we show the 1D rung 1 equation SISSO produced in Equation \ref{eq:maxpfc} (rung 1 means that the mathematical operators are only applied once to the primary features). The simplest one-dimensional descriptor found by SISSO for "$\mathrm{max\_pfc}$", $\mathrm{\left( \frac{bwdf_{at\_dist0}}{dist_{at\_neg\_bwdf0}} \right)}$, is the ratio of the strongest bond $\mathrm{(bwdf_{at\_dist0})}$ to the corresponding bond length $\mathrm{(dist_{at\_neg\_bwdf0})}$, identified by the bond-weighted distribution function (BWDF).\cite{deringer_bonding_2014} The found descriptor has a Pearson correlation coefficient of -0.91 with "$\mathrm{max\_pfc}$". This highly interpretable descriptor is in line with previous work highlighting the relation between bond lengths  \cite{bagheri2023identification, zahn2025experimental} and bonding strengths \cite{deringer2015vibrational, hempelmann2022orbital} and projected force constants. 

\begin{equation}
\mathrm{max\_pfc = 1.77 - 4.25 \left( \frac{bwdf_{at\_dist0}}{dist_{at\_neg\_bwdf0}} \right)}
\label{eq:maxpfc}
\end{equation}

Importantly, the resulting expression is highly interpretable and consistent with physical intuition: shorter, stronger bonds generally give rise to larger force constants. The rediscovery of this relationship increases confidence in the overall workflow and suggests that bonding descriptors capture physically and chemically meaningful and transferable aspects of chemical bonds across diverse materials.

The second example is the lattice thermal conductivity at 300 K ($\mathrm{log\_klat\_300}$), a directly measurable property that is central to applications ranging from thermoelectrics to thermal management. In this case, SISSO again identifies a descriptor containing a bonding term derived from the BWDF, namely its skewness, together with the atomic density:

\begin{equation}
\mathrm{log\_klat\_300 = -0.41 - 2.9 \left( \frac{pair_{bwdf\_skew\_mean}}{DensityFeatures_{vpa}} \right)}
\label{eq:logklat}
\end{equation}

The skewness of the BWDF can be interpreted as a measure of bond heterogeneity, reflecting the coexistence of strong covalent bonds alongside predominantly ionic ones. Bond heterogeneity has previously been identified as a key design principle for low thermal conductivity materials.\cite{pal2018bonding, sato2021bonding, jin2022bonding, powell2025exploiting} The second term in the descriptor, the atomic density, is likewise physically meaningful and consistent with established models of thermal transport.\cite{slack_thermal_1964, slack_nonmetallic_1973,cahill_lower_1992,agne_minimum_2018, bustamante2026thermal}

Unlike the established models, which predominantly rely on phonon and elastic descriptors such as Debye temperatures, sound velocities, or Grüneisen parameters, the descriptor identified here directly incorporates information about the underlying bonding landscape. While quantities derived from harmonic phonon theory are highly predictive, they are often less useful as design variables because they are not directly accessible through synthetic control. In contrast, bond heterogeneity can be translated into a chemically intuitive design principle. The recovery of this relationship through symbolic regression therefore demonstrates that bonding data can yield both predictive models and chemically meaningful guidance for materials design.

The examples above highlight that the inclusion of bonding data not only improves ML predictions but, when combined with techniques such as symbolic regression, also enables the construction of highly chemically intuitive models. Bonding descriptors are rooted in a rich and longstanding body of chemical theory and literature, allowing model outputs to be interpreted through familiar concepts and established mechanisms. Beyond improving model explainability, this connection enables the data-driven extraction of chemically meaningful design rules to guide materials discovery.

Looking forward, combining foundation models for predicting bonding properties, symbolic regression or other interpretable ML models, and self-driving laboratories could establish a powerful closed-loop framework for scientific discovery, particularly when targeting directly measurable properties. In such a workflow, AI systems use existing (bonding) data to generate hypotheses in the form of design rules, which are then used to propose candidate materials and experiments. Experimental results then refine the underlying hypotheses and guide the next round of measurements. By integrating autonomous design-rule generation into the materials discovery loop, self-driving labs could move beyond optimizing material properties toward uncovering the governing principles that control them. Self-driving laboratories for interpretable scientific discovery have recently been implemented; however, without bonding-based hypotheses.\cite{desai_autoscilab_nodate}

Extracted design rules could also provide a foundation for more trustworthy autonomous materials discovery workflows. Because these rules are expressed in terms of chemically meaningful bonding characteristics, AI-generated hypotheses can be rationalized, scrutinized, and validated before experimental resources are committed. Such chemically grounded explanations may become increasingly important as self-driving laboratories move toward autonomous decision-making.

More broadly, chemical bonding offers a representation that is naturally aligned with the language chemists use to describe materials. Concepts that frequently appear throughout the literature, such as strong covalent bonding, multicenter interactions, or bond heterogeneity, could potentially be mapped onto quantitative bonding descriptors and, conversely, translated back into human-interpretable chemical concepts. Bonding foundation models may therefore provide a bridge between machine-learned representations, electronic-structure theory, and the accumulated qualitative knowledge encoded in decades of scientific publications. 

In this view, chemical bonding is not merely a set of descriptors for property prediction, but a compact and chemically meaningful representation through which materials can be understood, predicted, and ultimately designed.

\subsection{Synthesizability predictions}

Synthesizability of solids remains a major challenge in the computer-aided discovery of materials with targeted properties. We begin by outlining this challenge and explaining why concepts from chemical bonding may help address it.

Whether a material can be synthesized depends on thermodynamic stability, kinetic accessibility, and synthesis conditions. While synthesizability is hard to predict for both molecules and solids, the problem is particularly severe for solids because the relevant thermodynamic and kinetic processes often require simulations of extended systems, as diffusion, nucleation, defects, phase transformations, and interface reactions play major roles.\cite{kumar_solid-state_2022} 

As a result, synthesizability assessments for solids often rely on computing the energy above the convex hull, a purely thermodynamic descriptor with well-known limitations.\cite{cavignac_ai-driven_2026} Such metrics are defined relative to the set of currently known phases at similar compositions and are therefore often inherently incomplete. Furthermore, relating energies above hull to experimental accessibility depends on additional heuristics (e.g., amorphous limits\cite{aykol_thermodynamic_2018}). Even finite-temperature effects are often neglected in convex-hull-based assessments.\cite{benitez_thermodynamics-informed_2026} 

The approaches described so far focus primarily on thermodynamic stability and provide little information about kinetic accessibility. In recent years, increasing efforts have been devoted to incorporating kinetic factors into synthesizability assessments. For example, explicit nucleation models combined with first-principles information have been employed to account for nucleation kinetics,\cite{he_precursor_2023} while cellular automaton frameworks combining first-principles stability and transport information have been used to model solid-state reactions.\cite{karan_ion_2026}

Alternatively, researchers have developed purely data-driven approaches for both synthesizability prediction and identifying potential synthesis pathways. One class of methods relies on positive-unlabeled (PU) learning models trained on experimentally realized materials together with large sets of unlabeled candidate compounds represented by their compositions or crystal structures.\cite{jang_structure-based_2020, jang_synthesizability_2024} For crystal structures, these approaches often rely on graph-based representations. While powerful, these models are biased toward assigning higher synthesizability scores to materials that are structurally and chemically similar to those present in the training database.\cite{kim_enhancing_2026} This limitation can be alleviated by incorporating complementary information that structure-based graph-based representations do not capture.\cite{kim_enhancing_2026} Another strategy is co-training, which can partially overcome biases associated with individual representations.\cite{amariamir_syncotrain_2025}

Another class of data-driven approaches relies on literature-mined datasets of chemical reactions. Advances in natural language processing and large language models have enabled the extraction of synthesis recipes from the literature, which are increasingly used to suggest synthesis pathways through machine-learning approaches.\cite{kononova_text-mined_2019, he_precursor_2023,prein_retro-rank-_2025} Typically, these datasets include reactant and product compositions, along with process descriptions and synthesis operations that can be used for machine-learning models. Structural information is typically neglected or only incorporated indirectly (e.g., by associating compositions with their most stable known polymorph).\cite{prein_retro-rank-_2025}

Importantly, most of these approaches do not explicitly encode quantum-chemical bonding descriptors. While some representations incorporate structural descriptors related to bonding, such as coordination environments, they often do not include bonding descriptors derived directly from the electronic structure. Yet chemists routinely rely on (quantum-chemical) bonding concepts when assessing synthesizability, for example, to rationalize phase stability, transport properties, phase transitions, nucleation processes, and chemical reactivity. We therefore believe that incorporating such bonding representations, either directly at the graph level or through statistical descriptors, could introduce chemically meaningful information across these different synthesizability frameworks. 

Beyond direct prediction tasks, foundation models trained on bonding descriptors may provide new opportunities for knowledge extraction and hypothesis generation. In particular, a promising direction could be large-language-model-assisted generation of chemically interpretable rules based on bonding descriptors. A related approach has recently been demonstrated using, among other descriptors, structural descriptors related to bonding.\cite{choi_compiling_2026}

\subsection{Generative Models}

Generative models rely on representations of materials. While modern approaches have demonstrated impressive performance based on structural inputs only, they have been criticized for generating chemically unreasonable structures, as reported, for example, by \cite{John_Jelfs_Ganose_Giunchiglia_2026}. Further, the relationship between structural representations and many target properties remains indirect.\cite{de_breuck_generative_2025} Properties such as stability, charge transport, reactivity, and, to some extent, synthesizability ultimately arise from the electronic structure. Bonding descriptors can therefore be viewed as compact representations of chemically relevant aspects of the electronic structure that are directly connected to these properties, including the chemical soundness. This may be particularly advantageous for conditional generation, where the goal is not merely to generate valid crystal structures but to generate materials with targeted functionalities. Incorporating bonding information could therefore provide a representation space that is more closely aligned with the physical origins of the desired properties. 

To address the criticism that generative models produce chemically unreasonable structures, as reported in\cite{John_Jelfs_Ganose_Giunchiglia_2026}, the authors of the study added oxidation-state conditions to generate reasonable materials that obey charge-neutrality rules in an output layer. Bonding concepts could provide an alternative or complementary strategy. For example, bond-valence sum rules, coordination numbers, and local environments could be incorporated as descriptors or constraints to guide generation toward chemically plausible bonding networks and more realistic materials.

Several recent studies have explored augmenting structural representations with descriptors more directly connected to electronic structure and related to bonding concepts. For example, \cite{Park_Choi_Jung_2026} argued
that atomic representations alone may be insufficient for conditional material generation because properties such as band gap, magnetic density, stability, and ionic conductivity ultimately depend on electronic structure. This motivates incorporating electronic-structure features into generative models' inputs to better represent materials and later conditionally generate new materials with the desirable properties. To this end, the authors developed ChargeDIFF,\cite{Park_Choi_Jung_2026} which incorporates electron charge density features as voxel-based data alongside the atomic representation. This joint representation improved the conditional generation of new materials, including those with properties such as band gap, crystal density, magnetic density, and Li-ion battery cathode materials, compared with the model's atomic-only representation. Moreover, MatterGen-e$^-$, \cite{Okuda_Takahara_Mizoguchi_2026} an enhanced diffusion model built on MatterGen,\cite{Zeni_Pinsler_Zugner_Fowler_Horton_Fu_Wang_Shysheya_Crabbe_Ueda} incorporates site-resolved (node) features such as Bader charges and the density of states (DOS) (similar to our ESF-MEGNet model) to train the diffusion models conditioned on formation energy and band gap. These joint models generate new materials that better match target properties and increase the likelihood of meeting 
validity, stability, uniqueness, and novelty (VSUN) criteria. 

While these approaches incorporate explicit electronic-structure descriptors, a natural question is whether similar information is already encoded in the latent representations learned by foundation models. Some of us have recently incorporated representations learned by pretrained MLIPs into a materials generative framework.\cite{hagemann_representations_2026} This approach is based on the assumption that MLIPs trained on energies and forces must implicitly encode information about chemical bonding and local atomic environments that govern stability within their latent representations. These representations can then be exploited as coarse coordinates for materials generation. Interestingly, the resulting generative model produced higher-quality structures, particularly with respect to stability, suggesting that MLIP representations capture chemically relevant information beyond purely geometric descriptions. This observation further supports the idea that representations learned by foundation models contain information closely related to chemical bonding and may provide a promising foundation for future bonding-aware generative models. 

If latent representations from MLIPs already improve materials generation, we can ask whether explicitly incorporating chemically interpretable bonding descriptors into their representation, or into those of related graph-based ML models, could provide additional benefits. We hypothesize that adding bonding properties to these representations such as ICOHP and ICOBI, along with enforcing chemical rules such as bond valence sum, charge neutrality, and coordination, could also help unlock the potential of conditional generative models, enabling the models to generate materials that better match desirable, targeted material properties. These bonding features could augment and enrich the trained model's latent representation. In this regard, these additional features may explain, connect, or describe filled regions or empty gaps in the diffusion model's latent space, later guiding the model to generate VSUN crystalline and disordered materials, especially for material properties highly correlated with the added features.

\subsection{Potential of Joint-Embedding Predictive Learning Across Chemical Bonding Analyses}\label{sec:JEPA}

Quantum-chemical bonding analysis methods aim to translate electronic-structure calculations into chemically interpretable quantities. However, no individual bonding descriptor provides a complete description of the underlying electronic structure, as each emphasizes different aspects of chemical bonding. In section \ref{sec:bonding_amb}, we discussed the case where geometric and quantum-chemical bonding algorithms do not fully agree on bonding detection in an intermetallic compound. In section \ref{sec:limits_comp_mater_representation}, we discussed the underlying sources of ambiguities for quantum-chemical bonding descriptors, which are not merely algorithmic. As quantum mechanics does not uniquely determine how electrons should be partitioned into atoms, bonds, or localized orbitals, different quantum-chemical bonding descriptions can be used for the same electronic structure. Thus, there is no reason to believe that any individual bonding analysis will provide a complete or unique description of chemical bonding.

Rather than treating this non-uniqueness as a limitation, we suggest using it as a source of information. The goal could therefore be not to identify a single best bonding description, but to learn which aspects of different bonding analyses are shared, complementary, or method-specific. Assuming different methods emphasize different aspects of the same underlying electronic structure, their relationships, agreements, and disagreements may define a richer representation for materials property prediction and help identify complex bonding situations. Joint-embedding predictive architectures (JEPAs)\cite{lecun2022path, dawid2024introduction} offer a natural framework for learning such representations. A trained JEPA encoder learns to predict the latent representation of a complex target from a given context, also known as views. In this process, it discards fine details that are difficult or impossible to predict while compressing them into a representation that retains shared information. This framework seems well-suited to our bonding analysis data, as each analysis carries method-specific artifacts, such as the geometric algorithm used, density-partitioning choices, and dependence on the local basis used in projection, that a reconstruction objective would force a model to reproduce. JEPA-based learning is already being explored for molecules and crystalline materials.\cite{piccoli2026joint,petersen2026atomjepajointembeddingpredictivearchitecture, rottach2026moljepamultimodaljointembedding, liu2026crysjepaacceleratingcrystaldiscovery,ramlaoui2026triforcesaugmentingatomisticgnns} These proposed approaches learn mainly from structures, energies, or molecular data; to our knowledge, none treats several bonding analyses of the same material as complementary views.



However, realizing this would require a dataset in which the same materials are analyzed with several bonding methods under consistent calculation settings and levels of theory, such that differences between analyses reflect the methods rather than the inputs. For each material, the crystal structure, the electronic-structure output, and each bonding analysis result would form separate views of the same compound. These views can also take different forms, for example: real-space descriptors such as the electron localization function (ELF) are three-dimensional fields that can be treated like images, while projected-orbital analyses give energy-resolved COHP/COBI spectra and population-based schemes give scalars. Training would then be a prediction task: one view is hidden, and the model must predict a latent representation of it from the remaining views. Information that can be predicted from the other views is shared across methods and therefore robust. Information that can be predicted only from some views is complementary. Information that no other view can predict is specific to that method. Comparing how well each view is predicted from the others therefore turns the question of which information is robust, complementary, or method-dependent into something measurable. The trained encoder would then serve as a pretrained bonding representation for downstream property prediction, in a similar way as such self-supervised models are used in other works.\cite{petersen2026atomjepajointembeddingpredictivearchitecture, rottach2026moljepamultimodaljointembedding}

The pretrained representation should predict properties not seen in training, such as phase stability, elastic and vibrational response, or defect formation, better than descriptors from any single analysis, particularly for unseen chemistries. It must also outperform a representation learned from the crystal structure alone. Because the bonding analyses are computed from the structure, only this comparison shows that they add information. We note, however, that the resulting JEPA representation would likely be neither low-dimensional nor easily interpretable. As such, it should not be regarded as an ideal bonding descriptor for chemical reasoning. Instead, its primary value would lie in predictive performance and as a unified representation of bonding information.

We envision two further uses that follow from having such a pretrained JEPA model. First, the structure encoder learns to predict the bonding views, so it can provide bonding-related information. Thus, it needs only a crystal structure, so it could supply bonding-informed features to machine-learned interatomic potentials without running the bonding analyses on every new structure. Second, how well each view can be predicted from the others shows which analyses carry overlapping information and which carry unique information. When bonding analysis methods disagree, for instance, in the case of electron-rich versus electron-deficient assignments for GeTe-type compounds,\cite{jones2023myth,muller2024chemical,wuttig2023revisiting,gatti2025comment, wuttig2024metavalent} or the example presented in section \ref{sec:bonding_amb}, such a model could identify compounds that warrant closer chemical investigation of a potentially more complex bonding situation.

Beyond JEPA models that integrate bonding information derived from geometric and quantum-chemical data, textual descriptions of materials could also be incorporated. Previous work has demonstrated that crystal-structure representations can be connected with language models through synthetic property narratives.\cite{park_scaling_2026}  In a similar spirit, chemical bonding may provide a common representation layer between atomistic data and chemical language. As a compact and chemically meaningful representation of electronic structure, bonding could enable multimodal models that jointly learn from structural, electronic, and textual descriptions of materials.

\subsection{Toward Next-Generation Bonding Representations and Foundation Models}

The strong correlations between latent representations of foundation machine-learning interatomic potentials (MLIPs) and quantum-chemical bonding descriptors suggest that both quantities can be learned jointly within a single model. Rather than treating bonding analyses as a separate step, future MLIP architectures could employ dedicated prediction heads for quantities such as ICOHPs, ICOBIs, and related bonding descriptors. The edge-centric design of architectures such as Allegro appears particularly well suited for this purpose. Such models could simultaneously provide energies, forces, and chemically interpretable bonding information. This would not only facilitate the analysis of chemical trends but also provide direct insight into failure modes of MLIPs, as unusual bonding patterns could be identified and traced back to specific local environments. Furthermore, one could use bonding descriptors such as ICOHPs and ICOBIs as additional training targets. By requiring a model to reproduce not only energies and forces but also chemically meaningful bond strengths and bond orders, the latent representation could be guided toward chemically consistent solutions. Such a framework would naturally combine predictive accuracy with chemical interpretability and provide a direct route toward chemistry-aware foundation models. Further, they would significantly reduce the black-box character of foundation MLIPs by providing access to chemically interpretable targets. Nevertheless, the benefits and drawbacks of using bonding descriptors have to be investigated and understood in a similar way as the effect of symmetry is currently under debate.\cite{langer_probing_2024}

An important, largely unexplored aspect is how bonding descriptors for materials depend on the level of electronic-structure theory. The bonding databases discussed in this work are predominantly based on GGA-level calculations, which are generally adequate for structural and vibrational properties but often exhibit deficiencies for electronic properties. Future bonding foundation models could provide a particularly attractive route for transferring bonding analyses across levels of theory. Because pretrained MLIPs already encode substantial bonding information, only comparatively small datasets of energies and forces and/or bonding data may be required to fine-tune such models toward higher levels of theory. This would enable the rapid generation of chemically interpretable bonding descriptors at levels of theory that are currently prohibitively expensive for large-scale bonding databases. More broadly, it may open a path toward transfer learning from electronic-structure data to experimentally measured properties, helping to bridge the gap between theory and experiment. Fine-tuning foundation models on experimental data may alter the bonding information encoded in their latent representations. Tracking how bonding descriptors evolve during transfer learning could reveal which bonding concepts matter most for experimental observations. In turn, this could facilitate the development and validation of bonding descriptors that are directly predictive of experimentally measurable properties. More generally, it may even enable the discovery of entirely new compact and low-dimensional bonding descriptors.

As shown throughout this work, alternative low-dimensional representations of electronic structure may capture chemically relevant information for materials-property prediction more effectively than established descriptors such as ICOHP, ICOBI, or Electron localization functions (ELF). Thus, the question may shift from evaluating the predictive power of existing bonding descriptors toward identifying low-dimensional representations of the electronic structure that best capture those aspects governing materials properties and chemical reactivity. Such representations could also be discovered through data-driven optimization and symbolic-regression approaches directly from the density matrix and related quantities, or from the electron density itself, complementing descriptors originally developed from chemical intuition and reasoning. Importantly, predictive performance alone should not be the sole criterion. Complexity\cite{vladislavleva_order_2009} and the dimensionality of the resulting representation are equally important, as they determine the extent to which the descriptor provides chemical insight rather than merely predictive accuracy.

\section{Conclusion}

The results presented in this work suggest that the relationship between chemical bonding and machine learning is not one of replacement but of mutual benefit. Thus, returning to our title question, we still indeed need the concept of chemical bonds. One consequence of the age of big data and AI is a shift in perspective: rather than asking which bonding picture is the most "correct", we can increasingly ask how useful a bonding descriptor is for explaining, predicting, and designing materials. This is especially useful for the inherently ambiguous concept of chemical bonding, even when defined quantum-mechanically.  Nevertheless, existing bonding descriptors remain valuable because they provide chemically interpretable and often complementary information that can improve predictive performance beyond purely distance-based representations. At the same time, the strong correlations between quantum-chemical bonding descriptors and the latent representations of foundation machine-learned interatomic potentials indicate that these models learn chemically meaningful information even without explicit supervision. This suggests that traditional descriptors such as ICOHPs and ICOBIs may not mark the endpoint of bonding analysis, but instead serve as interpretable low-dimensional projections of richer representations learned by machine-learning models. Looking forward, machine learning offers an opportunity not only to quantitatively evaluate existing bonding concepts, but also to discover new low-dimensional electronic-structure representations optimized for explaining, predicting, and ultimately designing materials. This could include predicting experimentally relevant properties such as transport and mechanical properties, as well as synthesizability. More generally, we propose that chemical bonding can be understood as a learnable representation of electronic structure whose usefulness can be evaluated quantitatively through predictive power, dimensionality, and complexity. Furthermore, chemical bonding is the fundamental language through which chemistry rationalizes materials and their properties. If a materials genome, or something close to it, exists, it should therefore be possible to encode it through a representation of chemical bonding. Conversely, if no bonding representation can capture the information necessary to explain, predict, and design materials behavior across chemistry, our current understanding of the chemical bond may require revision.

\section{Methods}

\phantomsection
\addcontentsline{toc}{subsection}{\texorpdfstring{\ce{LaRu2P2} and \ce{CeAl2Ga2} Calculations}{LaRu2P2 and CeAl2Ga2 Calculations}}
\subsection*{\ce{LaRu2P2} and \ce{CeAl2Ga2} Calculations}

We used atomate2 v0.0.22\cite{ganose2025atomate2} for the DFT calculations and bonding analyses of \ce{CeAl2Ga2} and \ce{LaRu2P2}. 
We conducted the DFT calculations with the Vienna Ab-Initio Simulation Package (VASP, v6.4.3)\cite{Kresse_Furthmüller_1996_1, Kresse_Furthmüller_1996_2, Kresse_Joubert_1999} and the bonding analyses with LOBSTER v5.1.1.\cite{deringer2011crystal, maintz2013analytic, maintz2016lobster,nelsonLOBSTERLocalOrbital2020}
To simulate the pressure dependence of \ce{LaRu2P2}, we relaxed the compound at various fixed volumes using the revised version of the Perdew-Burke-Ernzerhof (PBEsol) exchange-correlation functional\cite{perdew_pbesol_2008} and a kinetic energy cutoff of 680\,eV for the plane-wave basis. Energy and force convergence criteria of 1E-6\,eV and 0.02\,eV/{\AA} were applied.
We then recomputed the relaxed structures using the Perdew-Burke-Ernzerhof (PBE) exchange-correlation functional\cite{perdew_pbe_1996} and a kinetic energy cutoff of 520\,eV for the plane-wave basis, and projected onto a local basis using LOBSTER. 
The \ce{CeAl2Ga2} DFT calculations used PBE and a kinetic energy cutoff of 520\,eV as well. For postprocessing and coordination analyses, pymatgen v2025.6.14 was used.

\phantomsection
\addcontentsline{toc}{subsection}{Allegro \& MACE Analysis}
\subsection*{Allergo \& MACE Analysis}

We loaded the pretrained Allegro model "Allegro-OAM-L-0.1" up to the "per\_type\_energy\_scale\_shift" processing layer. Then extracted E$_{ij}$ and 1,024-dimensional $x_{ij}$ from the nequip\cite{Batzner_Musaelian_Sun_Geiger_Mailoa_Kornbluth_Molinari_Smidt_Kozinsky_2022} `AtomicDataDict` keys 'edge\_energy' and 'edge\_features', respectively, from the transformed ASE\cite{Hjorth_Larsen_ASE} atoms. 
We generated the MACE node-level descriptors using the function 'get\_descriptors' from the 'MACECalculator' class that loaded the model 'mace-mpa-0-medium.model'. The 'get\_descriptors' function extracts the node features of 2 interaction blocks, accounting for 256 node embeddings (i.e., 128 embeddings from each block). We perform PCA on the normalized features from both MLIP models to visualize their respective latent representations.

To compute $\mathrm{dCor}$, we used the $\mathrm{dCor}$\cite{Ramos_dcor_2023} Python package, version 0.7. $\mathrm{dCor}$\cite{Szekely_Rizzo_Bakirov_2007} is defined by the equation below: 
\begin{equation}
\mathcal{R}^2(X,Y) = 
\frac{\mathcal{V}^2(X,Y)}
{\sqrt{\mathcal{V}^2(X,X)\,\mathcal{V}^2(Y,Y)}}
\label{eq:DCOR_eq}
\end{equation}
\begin{equation} \operatorname{dCor}(X,Y) = \sqrt{\mathcal{R}^2(X,Y)}. \label{eq:dcor} \end{equation}
where $\mathcal{V}^2(X,Y)$ is the distance covariance, and $\mathcal{V}^2(X,X), \mathcal{V}^2(Y,Y)$ are distance variances. 

We also define the coefficient of determination ($R{^2}$) as: 
\begin{equation}
R^2=1-\frac{\sum_{i=1}^{n}(y_i-\hat{y}_i)^2}
{\sum_{i=1}^{n}(y_i-\bar{y})^2}
\label{eq:r2}
\end{equation}
where $y_i$ are the actual values. $\bar{y}$ is the mean of all the actual values. $\hat{y}_i$ are the predicted values.

\phantomsection
\addcontentsline{toc}{subsection}{ESF-MEGNet Variants}
\subsection*{ESF-MEGNet Variants}

To train GNN models, we use several datasets for the various trained properties. The materials in the datasets come from the Materials Project,\cite{Jain_Ong_Hautier_Chen_Richards_Dacek_Cholia_Gunter_Skinner_Ceder_et_al_2013} and we use the Application Programming Interface (API)\cite{Ong_Cholia_Jain_Brafman_Gunter_Ceder_Persson_2015} to obtain some of the material property datasets. We obtain datasets of 1,153 materials with average MSD and 1,259 materials with the last phonon peak frequency from the phonon dataset in \cite{Petretto_Dwaraknath_P_C_Miranda_Winston_Giantomassi_Van_Setten_Gonze_Persson_Hautier_Rignanese_2018}, processed as conducted in \cite{Naik_Dhamrait_Ueltzen_Ertural_Benner_Rignanese_George_2026}. We also obtain a dataset of bulk and shear moduli for 3,073 materials from \cite{De_Jong_Chen_Angsten_Jain_Notestine_Gamst_Sluiter_Krishna_Ande_Van_Der_Zwaag_Plata_et_al_2015}. We process all the above datasets similarly to \cite{Naik_Dhamrait_Ueltzen_Ertural_Benner_Rignanese_George_2026}. We constructed our Vickers Hardness dataset, which consists of 3,072 materials, using Tian's macroscopic description of hardness \cite{Tian_Xu_Zhao_2012} (see equation~\ref{eq:Hv}), which modifies Chen’s formula.\cite{Chen_Niu_Li_Li_2011} We used Tian's model due to its good agreement with experimental results,\cite{Tian_Xu_Zhao_2012, Zhao_Cui_Wang_Xu_Liu_Yu_He_Zhou_Wang_Tian_2010, Li_Xue_2009, Guo_Li_Liu_Yu_He_Liu_Xu_Tian_Wang_2008, Simunek_Vackar_2006, Gao_He_Wu_Liu_Yu_Li_Zhang_Tian_2003} and the model has been implemented in high-throughput and machine learning studies \cite{Chen_Schmidt_Yan_Vohra_Chen_2021, Al_Fahdi_Rodriguez_Ouyang_Hu_2021, Al_Fahdi_Ouyang_Hu_2021, Ojih_Al_Fahdi_Rodriguez_Choudhary_Hu_2022, Mukherjee_Sahu_Losego_Gutekunst_Ramprasad_2024} as well.

\begin{equation}
H_V = 0.92\,k^{1.137}G^{0.708} = 0.92\,(G/K)^{1.137}G^{0.708}
\label{eq:Hv}
\end{equation}
where $G$ is the shear modulus and $k$ (lowercase) is Pugh's ratio, defined as $k=G/K$, where $K$ (uppercase) is the bulk modulus. Further, we define the baseline MEGNet and ESF-MEGNet model variants' performance metric, MAE, as follows:
\begin{equation}
\mathrm{MAE}=\frac{1}{n}\sum_{i=1}^{n}\left|y_i-\hat{y}_i\right|
\label{eq:mae}
\end{equation}
where $y_i$ are the actual properties. $\hat{y}_i$ are the predicted properties. We also compute the MAE Improvement (\%) as shown by equation \ref{eq:mae_improvement} below:
\begin{equation} \mathrm{MAE\,Improvement}~(\%) = \frac{\mathrm{MAE}_{\mathrm{MEGNet}}-\mathrm{MAE}_{\mathrm{ESF\text{-}MEGNet}}} {\mathrm{MAE}_{\mathrm{MEGNet}}} \times 100, \label{eq:mae_improvement} \end{equation}
As shown in the equation, we compute the MAE Improvement (\%) for ESF-MEGNet relative to the baseline MEGNet model. We also detail the training hyperparameters and model architecture parameters for all MEGNet variants in Table~\ref{tab:training_hyperparameters} and Table~\ref{tab:model_parameters}, respectively.

\
\begin{table}[pos=h] 
\centering 
\caption{Training hyperparameters used in all ESF-MEGNet variants.} \begin{tabular}{ll} 
\hline \textbf{Hyperparameter} & \textbf{Value} \\ 
\hline Optimizer & Adam \\ Learning rate & $1\times10^{-3}$ \\ Scheduler & CosineAnnealingLR \\ Decay steps & 1000 \\ Batch size & 64 \\ Maximum epochs & 500 \\ Loss function & MAE \\ Evaluation metrics & MAE, RMSE, $R^2$ \\ Random seed & 42 (data split, model initialization, and feature generation) \\ Hardware & GPU \\ 
\hline 
\end{tabular} 
\label{tab:training_hyperparameters} 
\end{table}

\begin{table}[pos=ht]
\centering
\caption{Model architecture and parameters. We also refer the reader to Fig. 2 in the original MEGNet work,\cite{chen_graph_2019}, which explains how these dimensions change through the architecture. We used the same argument names in the code as the MEGNet model in MatGL v2.0.0.}
\begin{tabular}{ll}
\hline
Parameter & Value \\
\hline
Node embedding dimension & 118 \\
Edge embedding dimension & 50 \\
State embedding dimension & 2 \\
Number of convolution blocks ($n_{\mathrm{blocks}}$) & 3 \\
Input hidden layer sizes & (64, 32) \\
Convolution hidden layer sizes & (64, 64, 32) \\
Output hidden layer sizes & (32, 16) \\
Set2Set layers & 1 \\
Set2Set iterations & 2 \\
Activation function & Softplus2. The edge and node fusion branches used SiLU \\
Classification mode & False (regression) \\
State attributes included & True \\
Dropout rate & 0.0 \\
Cutoff radius (\AA) & 5.0 \\
Gaussian standard deviation (\AA) & 1.0 \\
\hline
\end{tabular}
\label{tab:model_parameters}
\end{table}

\phantomsection
\addcontentsline{toc}{subsection}{ESF-MEGNet Variants}
\subsection*{Software}

For the bonding analysis of \ce{LaRu2P2} and \ce{CeAl2Ga2}, we used atomate2 0.0.22, VASP 6.4.3, LOBSTER 5.1.1, pymatgen 2025.6.14.  

For the Allegro and MACE MLIP results, the following software was used: torch 2.4.0, ase 3.29.0, scikit-learn 1.9.0, pandas 3.0.6, numpy 2.5.3, matplotlib 3.11.1, pymatgen 2026.5.4, lobsterpy 0.6.1, mace-torch 0.3.16, cuequivariance-torch 0.11.1, seaborn 0.13.2, nequip-allegro 0.8.3, DGL 2.4.0+cu124.

For the ESF-MEGNet results, the following packages were used:  torch 2.4.0+cu121, dgl 1.1.3+cu121, lightning 2.4.0, torchmetrics 1.4.2, numpy 1.26.4, pandas 2.2.3, scipy 1.14.1, scikit-learn 1.5.2, ase 3.23.0, pymatgen 2024.10.3, matplotlib 3.11.2, tqdm 4.70.1, requests 2.34.2

\section{Data Availability}
The code and data to reproduce the results in the manuscript are available via \href{https://doi.org/10.5281/zenodo.23808718}{https://doi.org/10.5281/zenodo.23808718}.

\section{Acknowledgements}

M. A.-F.,  A. A. N., L. W., T.D., and J.G. were supported by ERC Grant MultiBonds (grant agreement no. 101161771; funded by the European Union. Views and opinions expressed are, however, those of the author(s) only and do not necessarily reflect those of the European Union or the European Research Council Executive Agency. Neither the European Union nor the granting authority can be held responsible for them.) The authors gratefully acknowledge the Gauss Centre for Supercomputing e.V. (www.gausscentre.eu) for providing generous computing time on the GCS Supercomputer SuperMUC-NG at the Leibniz Supercomputing Centre (www.lrz.de) under Project No. pn73da.

\section{Declaration of Generative AI Usage}

Generative AI (Microsoft Copilot and Claude Opus 4.8/5.5) was used in preparing the publication. We used it to receive feedback on early drafts of the paper and to develop code associated with this paper. We used GitHub Copilot to write the README.md, docstrings, and type hints for the Allegro and MACE analysis workflows, but we developed the Allegro and MACE workflows with minimal AI input and feedback. GPT-6 Astra was used to generate ML model training pipelines. It was also used for literature searches and inspired the overview figure based on the paper's content. Based on these layout ideas, the authors created a corresponding figure.
The authors reviewed all results, figures, and text and edited the text as needed. They take full responsibility for the content of the work.

\printcredits

\bibliographystyle{model1-num-names}

\bibliography{cas-refs}





\cleardoublepage
\pagestyle{plain}

\makeatletter
\renewcommand*{\theHsection}{SI.\arabic{section}}
\renewcommand*{\theHsubsection}{SI.\arabic{section}.\arabic{subsection}}
\makeatother
\setcounter{figure}{0} 
\setcounter{equation}{0} 
\setcounter{table}{0} 
\setcounter{page}{1} 
\setcounter{section}{0}
\renewcommand{\thesection}{S\arabic{section}} 
\renewcommand{\thesubsection}{\thesection.\arabic{subsection}}
\renewcommand{\thefigure}{S\arabic{figure}}
\setcounter{figure}{0}
\makeatletter
\renewcommand{\thetable}{S\@arabic\c@table} 
\makeatother
\makeatletter
\renewcommand{\theequation}{S\@arabic\c@equation} 
\makeatother

\phantomsection
\addcontentsline{toc}{section}{Supplementary Information} 

\begin{center}
    \LARGE \textbf{Supplementary Information}\\[1em] 
    \large \textbf{In the age of big data and AI: do we still need chemical bonds?}\\[0.5em] 
\end{center}

\stmauthors 
\vspace{0.5em}
\noindent \small $^{a }$\textit{Bundesanstalt für Materialforschung und -prüfung (BAM), Unter den Eichen 87, Berlin, 12205,  Germany}\\
\small $^{b }$\textit{Friedrich Schiller University Jena, Max-Wien-Platz, Jena, 07743, Germany}\\

\vspace{1.5em}

\begin{figure}[pos=h]
    \centering
    \includegraphics[width=0.9\textwidth]{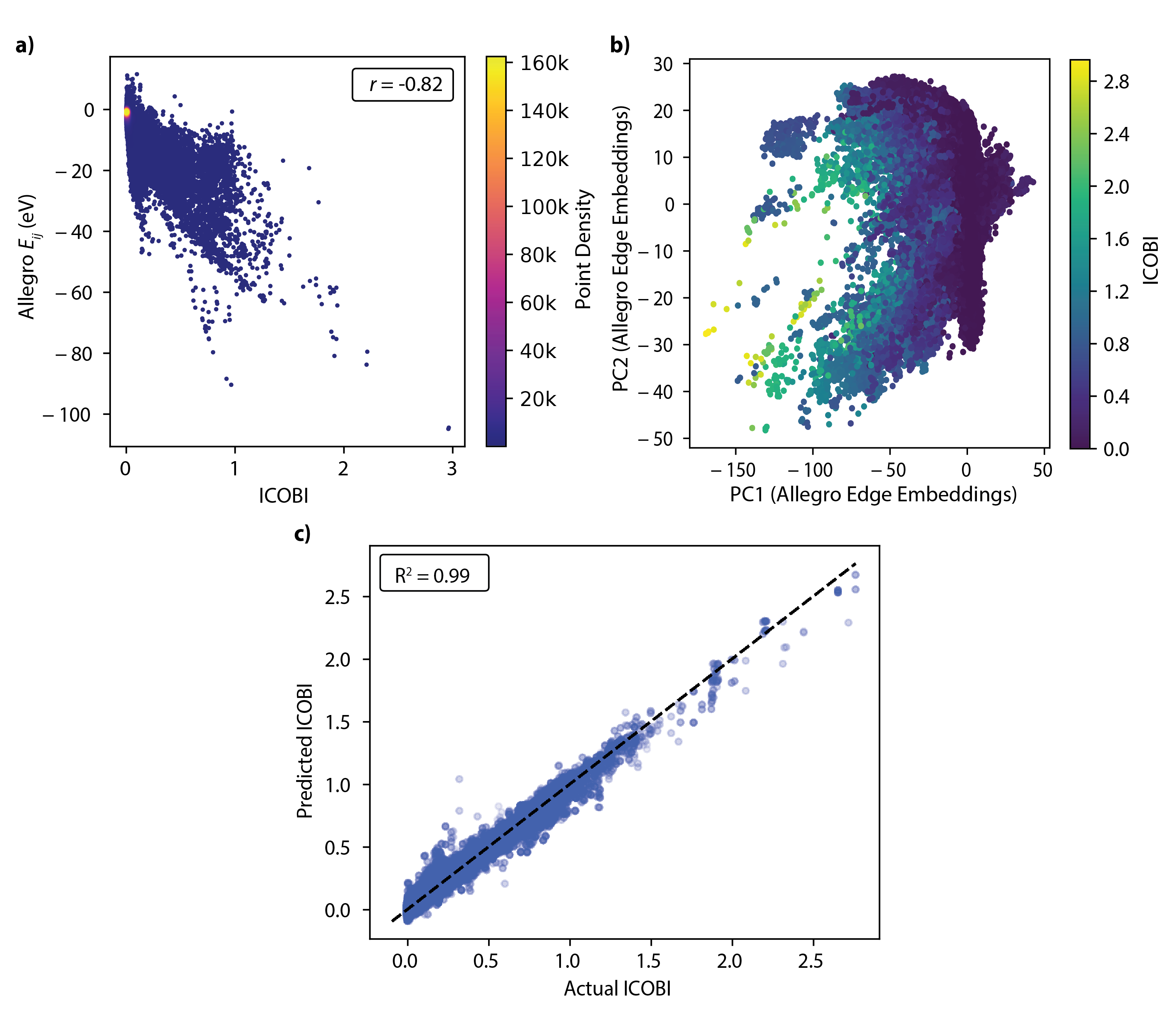}
    \caption{Investigation of Allegro Edge-wise energies ($E_{ij}$) and their embeddings ($x_{ij}$) extracted from the pretrained model "Allegro-OAM-L-0.1" from a bonding perspective. Plot of $E_{ij}$ against a) ICOBI with r showing the Pearson correlation between them, color-coded with the density of bond samples. Principal Component Analysis (PCA) of $x_{ij}$ yielded a 2D map of $\mathrm{PC1}$ and $\mathrm{PC2}$ projections of $x_{ij}$, color-coded by b) ICOBI. c) Parity plot of true vs. predicted ICOBI using linear probing of Allegro's edge latent representations (i.e., $x_{ij}$). The high correlation indicates that Allegro's latent space encodes and implicitly learns such bonding information.
    }
    \label{fig:SI_allegro}
\end{figure}

\begin{figure}
    \centering
    \includegraphics[width=0.9\textwidth]{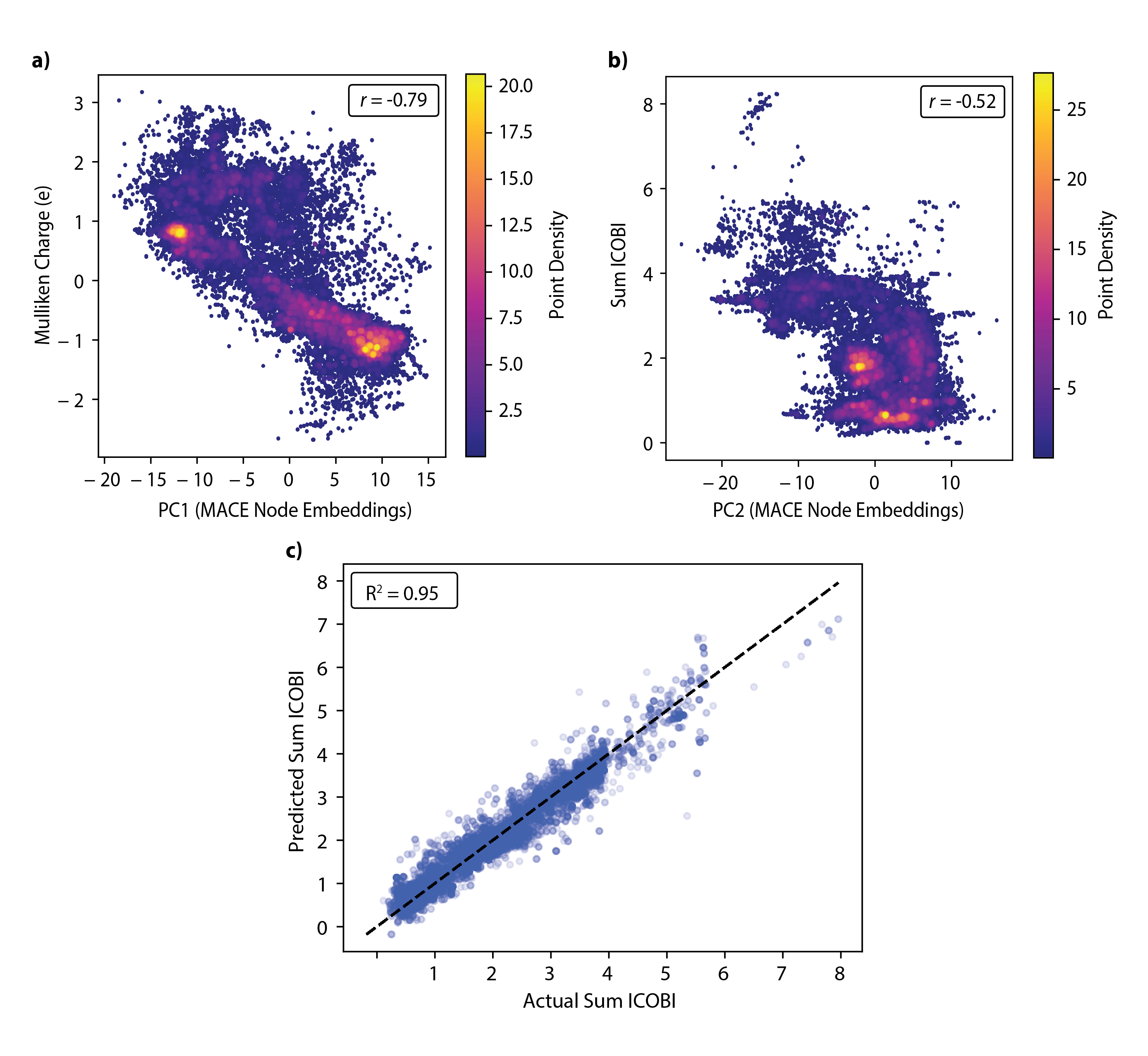}
    \caption{Investigation of the correlations between MACE latent representations and quantum-chemical bonding descriptors. a) Correlation between $\mathrm{PC1}$ and Mulliken Charge ($r = - 0.79$) b) Correlation between $\mathrm{PC2}$ and summed ICOBI ($r = -0.52$). c) Parity plot of predicted ICOBI by linear probing and actual ICOBI.}
    \label{fig:SI_mace}
\end{figure}

\begin{figure}
    \centering
    \includegraphics[width=0.9\textwidth]{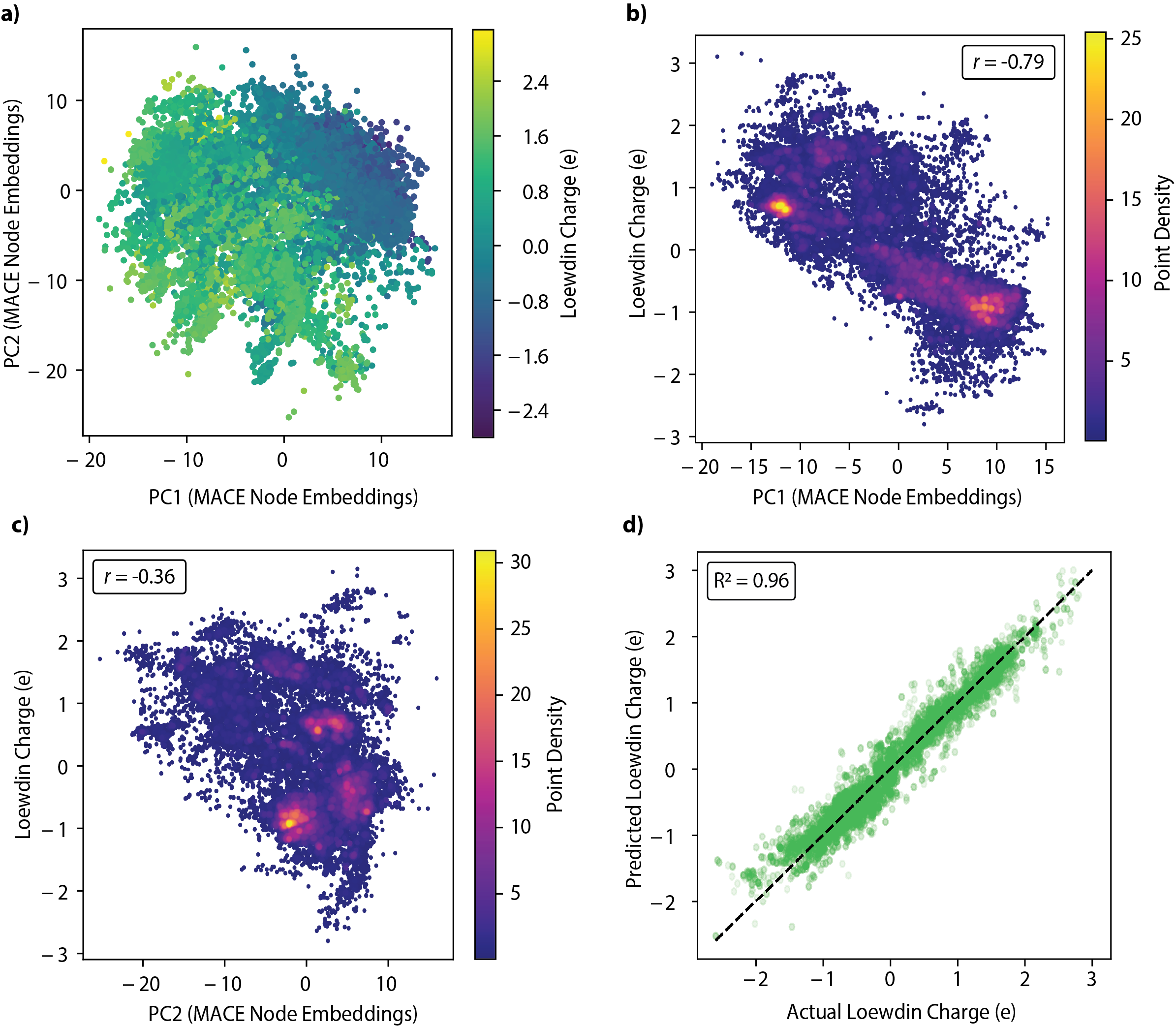}
    \caption{Löwdin analysis results. a) 2D map of PC projections on the MACE node embeddings color-coded by Löwdin charge, exhibiting a charge gradient aligned with PC1 (consistent with b). b) Correlation between PC1 and Löwdin charge, showing a strong negative relationship ($r = -0.79$). c) Correlation between PC2 and Löwdin charge, showing a strong negative relationship ($r = -0.36$). d) Parity plot of true vs. predicted Löwdin charge.}
    \label{fig:SI_mace_loew_results}
\end{figure}

\begin{table*} 
\centering 
\caption{Ten-fold cross-validation performance comparison for the last phonon peak. Mean ($\mu$) and standard deviation ($\sigma$) are reported across the ten folds. Values are shown to three decimal places for clarity, but all statistics are calculated from the full-precision results from the ten folds.} 
\label{tab:cv_results_last_peak} 
\scriptsize \begin{tabular}{llcccccccccccc} \toprule \textbf{Model} & \textbf{Metric} & \textbf{F1} & \textbf{F2} & \textbf{F3} & \textbf{F4} & \textbf{F5} & \textbf{F6} & \textbf{F7} & \textbf{F8} & \textbf{F9} & \textbf{F10} & \textbf{$\mu$} & \textbf{$\sigma$} \\ \midrule \multirow{2}{*}{ESF-MEGNet} & MAE & 32.558 & 37.655 & 28.581 & 36.691 & 27.622 & 40.759 & 36.899 & 37.762 & 30.798 & 27.958 & \textbf{33.728} & 4.795 \\ & $R^2$ & 0.988 & 0.986 & 0.986 & 0.978 & 0.988 & 0.984 & 0.961 & 0.979 & 0.980 & 0.989 & 0.982 & 0.008 \\ \midrule \multirow{2}{*}{ESF-MEGNet-rand} & MAE & 39.783 & 46.935 & 35.481 & 30.015 & 36.657 & 41.537 & 30.890 & 40.552 & 26.770 & 26.552 & 35.518 & 6.832 \\ & $R^2$ & 0.984 & 0.986 & 0.984 & 0.987 & 0.982 & 0.982 & 0.973 & 0.980 & 0.981 & 0.989 & \textbf{0.983} & \textbf{0.004} \\ \midrule \multirow{2}{*}{ESF-MEGNet-zero} & MAE & 33.438 & 47.141 & 30.521 & 34.584 & 33.864 & 44.287 & 35.222 & 52.024 & 28.565 & 30.846 & 37.049 & 7.917 \\ & $R^2$ & 0.987 & 0.986 & 0.984 & 0.979 & 0.988 & 0.982 & 0.970 & 0.965 & 0.976 & 0.987 & 0.980 & 0.008 \\ \midrule \multirow{2}{*}{MEGNet} & MAE & 52.894 & 34.053 & 30.551 & 46.650 & 33.751 & 42.527 & 32.756 & 37.308 & 29.600 & 29.166 & 36.926 & 7.967 \\ & $R^2$ & 0.960 & 0.991 & 0.982 & 0.969 & 0.982 & 0.984 & 0.976 & 0.985 & 0.977 & 0.989 & 0.980 & 0.009 \\ \bottomrule \end{tabular} \end{table*}

\begin{table*}
\centering 
\caption{Ten-fold cross-validation performance comparison for the mean-squared displacement at 300 K averaged over the non-
equivalent sites (Avg. MSD). Mean ($\mu$) and standard deviation ($\sigma$) are reported across the ten folds. Values are shown to three decimal places for clarity, but all statistics are calculated from the full-precision results from the ten folds.} 
\label{tab:cv_results_avgMSD} 
\scriptsize 
\begin{tabular}{llcccccccccccc} 
\toprule \textbf{Model} & \textbf{Metric} & \textbf{F1} & \textbf{F2} & \textbf{F3} & \textbf{F4} & \textbf{F5} & \textbf{F6} & \textbf{F7} & \textbf{F8} & \textbf{F9} & \textbf{F10} & \textbf{$\mu$} & \textbf{$\sigma$} \\ 
\midrule \multirow{2}{*}{ESF-MEGNet} & MAE & 0.072 & 0.075 & 0.082 & 0.066 & 0.077 & 0.078 & 0.069 & 0.077 & 0.073 & 0.082 & \textbf{0.075} & 0.005 \\ & $R^2$ & 0.906 & 0.889 & 0.895 & 0.894 & 0.894 & 0.877 & 0.901 & 0.812 & 0.872 & 0.850 & \textbf{0.879} & 0.029 \\ 
\midrule \multirow{2}{*}{ESF-MEGNet-rand} & MAE & 0.092 & 0.078 & 0.075 & 0.082 & 0.076 & 0.081 & 0.083 & 0.082 & 0.073 & 0.083 & 0.080 & 0.005 \\ & $R^2$ & 0.837 & 0.864 & 0.911 & 0.824 & 0.858 & 0.840 & 0.859 & 0.838 & 0.882 & 0.837 & 0.855 & 0.026 \\ 
\midrule \multirow{2}{*}{ESF-MEGNet-zero} & MAE & 0.082 & 0.081 & 0.075 & 0.076 & 0.082 & 0.085 & 0.091 & 0.081 & 0.068 & 0.087 & 0.081 & 0.007 \\ & $R^2$ & 0.856 & 0.853 & 0.908 & 0.825 & 0.841 & 0.835 & 0.810 & 0.820 & 0.908 & 0.823 & 0.848 & 0.035 \\ 
\midrule \multirow{2}{*}{MEGNet} & MAE & 0.073 & 0.084 & 0.085 & 0.072 & 0.082 & 0.087 & 0.078 & 0.083 & 0.066 & 0.074 & 0.078 & 0.007 \\ & $R^2$ & 0.910 & 0.863 & 0.889 & 0.883 & 0.839 & 0.826 & 0.884 & 0.807 & 0.908 & 0.861 & 0.867 & 0.034 \\ 
\bottomrule 
\end{tabular} 
\end{table*}

\begin{table*}
\centering 
\caption{Ten-fold cross-validation performance comparison for bulk modulus (K). Mean ($\mu$) and standard deviation ($\sigma$) are reported across the ten folds. Values are shown to three decimal places for clarity, but all statistics are calculated from the full-precision results from the ten folds.} 
\label{tab:cv_results_K} 
\scriptsize 
\begin{tabular}{llcccccccccccc} 
\toprule \textbf{Model} & \textbf{Metric} & \textbf{F1} & \textbf{F2} & \textbf{F3} & \textbf{F4} & \textbf{F5} & \textbf{F6} & \textbf{F7} & \textbf{F8} & \textbf{F9} & \textbf{F10} & \textbf{$\mu$} & \textbf{$\sigma$} \\ 
\midrule \multirow{2}{*}{ESF-MEGNet} & MAE & 0.072 & 0.070 & 0.082 & 0.088 & 0.086 & 0.091 & 0.076 & 0.084 & 0.074 & 0.087 & \textbf{0.081} & 0.007 \\ & $R^2$ & 0.909 & 0.910 & 0.904 & 0.790 & 0.863 & 0.703 & 0.910 & 0.879 & 0.911 & 0.893 & \textbf{0.867} & 0.069 \\ 
\midrule \multirow{2}{*}{ESF-MEGNet-rand} & MAE & 0.097 & 0.082 & 0.106 & 0.103 & 0.104 & 0.116 & 0.097 & 0.099 & 0.080 & 0.098 & 0.098 & 0.011 \\ & $R^2$ & 0.865 & 0.892 & 0.870 & 0.808 & 0.854 & 0.651 & 0.876 & 0.846 & 0.912 & 0.853 & 0.843 & 0.073 \\ 
\midrule \multirow{2}{*}{ESF-MEGNet-zero} & MAE & 0.091 & 0.100 & 0.099 & 0.106 & 0.101 & 0.116 & 0.102 & 0.087 & 0.082 & 0.091 & 0.097 & 0.010 \\ & $R^2$ & 0.872 & 0.847 & 0.825 & 0.808 & 0.845 & 0.621 & 0.862 & 0.874 & 0.891 & 0.869 & 0.832 & 0.078 \\ 
\midrule \multirow{2}{*}{MEGNet} & MAE & 0.091 & 0.083 & 0.095 & 0.099 & 0.106 & 0.103 & 0.097 & 0.100 & 0.076 & 0.096 & 0.095 & 0.009 \\ & $R^2$ & 0.853 & 0.874 & 0.861 & 0.785 & 0.805 & 0.676 & 0.874 & 0.843 & 0.908 & 0.850 & 0.833 & 0.065 \\ 
\bottomrule 
\end{tabular} 
\end{table*}

\begin{table*}
\centering 
\caption{Ten-fold cross-validation performance comparison for shear modulus (G). Mean ($\mu$) and standard deviation ($\sigma$) are reported across the ten folds. Values are shown to three decimal places for clarity, but all statistics are calculated from the full-precision results from the ten folds.} 
\label{tab:cv_results_G} 
\scriptsize 
\begin{tabular}{llcccccccccccc} 
\toprule \textbf{Model} & \textbf{Metric} & \textbf{F1} & \textbf{F2} & \textbf{F3} & \textbf{F4} & \textbf{F5} & \textbf{F6} & \textbf{F7} & \textbf{F8} & \textbf{F9} & \textbf{F10} & \textbf{$\mu$} & \textbf{$\sigma$} \\ 
\midrule \multirow{2}{*}{ESF-MEGNet} & MAE & 0.096 & 0.092 & 0.096 & 0.088 & 0.091 & 0.106 & 0.103 & 0.093 & 0.092 & 0.096 & \textbf{0.095} & 0.006 \\ & $R^2$ & 0.863 & 0.880 & 0.888 & 0.821 & 0.899 & 0.702 & 0.873 & 0.874 & 0.912 & 0.879 & \textbf{0.859} & 0.060 \\ 
\midrule \multirow{2}{*}{ESF-MEGNet-rand} & MAE & 0.113 & 0.123 & 0.113 & 0.117 & 0.121 & 0.154 & 0.125 & 0.121 & 0.097 & 0.118 & 0.120 & 0.014 \\ & $R^2$ & 0.829 & 0.831 & 0.850 & 0.802 & 0.811 & 0.595 & 0.831 & 0.862 & 0.908 & 0.810 & 0.813 & 0.083 \\ 
\midrule \multirow{2}{*}{ESF-MEGNet-zero} & MAE & 0.122 & 0.107 & 0.118 & 0.110 & 0.120 & 0.119 & 0.140 & 0.125 & 0.106 & 0.116 & 0.118 & 0.010 \\ & $R^2$ & 0.794 & 0.861 & 0.835 & 0.805 & 0.824 & 0.670 & 0.716 & 0.828 & 0.872 & 0.815 & 0.802 & 0.063 \\ 
\midrule \multirow{2}{*}{MEGNet} & MAE & 0.100 & 0.106 & 0.111 & 0.113 & 0.106 & 0.126 & 0.122 & 0.117 & 0.096 & 0.124 & 0.112 & 0.010 \\ & $R^2$ & 0.881 & 0.856 & 0.859 & 0.804 & 0.844 & 0.664 & 0.808 & 0.849 & 0.901 & 0.802 & 0.827 & 0.066 \\ 
\bottomrule 
\end{tabular} 
\end{table*}

\begin{table*}
\centering 
\caption{Ten-fold cross-validation performance comparison for Vickers hardness ($H_V$). Mean ($\mu$) and standard deviation ($\sigma$) are reported across the ten folds. Values are shown to three decimal places for clarity, but all statistics are calculated from the full-precision results from the ten folds.} 
\label{tab:cv_results_Hv} 
\scriptsize 
\begin{tabular}{llcccccccccccc} 
\toprule \textbf{Model} & \textbf{Metric} & \textbf{F1} & \textbf{F2} & \textbf{F3} & \textbf{F4} & \textbf{F5} & \textbf{F6} & \textbf{F7} & \textbf{F8} & \textbf{F9} & \textbf{F10} & \textbf{$\mu$} & \textbf{$\sigma$} \\ 
\midrule \multirow{2}{*}{ESF-MEGNet} & MAE & 0.130 & 0.118 & 0.120 & 0.112 & 0.135 & 0.117 & 0.119 & 0.122 & 0.117 & 0.126 & \textbf{0.122} & 0.007 \\ & $R^2$ & 0.757 & 0.780 & 0.801 & 0.782 & 0.674 & 0.776 & 0.790 & 0.803 & 0.788 & 0.696 & \textbf{0.765} & 0.044 \\ 
\midrule \multirow{2}{*}{ESF-MEGNet-rand} & MAE & 0.149 & 0.136 & 0.143 & 0.139 & 0.152 & 0.136 & 0.139 & 0.136 & 0.137 & 0.156 & 0.142 & 0.007 \\ & $R^2$ & 0.700 & 0.733 & 0.719 & 0.705 & 0.656 & 0.719 & 0.739 & 0.737 & 0.726 & 0.595 & 0.703 & 0.045 \\ 
\midrule \multirow{2}{*}{ESF-MEGNet-zero} & MAE & 0.140 & 0.127 & 0.140 & 0.136 & 0.148 & 0.134 & 0.136 & 0.137 & 0.138 & 0.137 & 0.137 & 0.005 \\ & $R^2$ & 0.715 & 0.775 & 0.737 & 0.716 & 0.675 & 0.730 & 0.748 & 0.743 & 0.742 & 0.686 & 0.727 & 0.030 \\ 
\midrule \multirow{2}{*}{MEGNet} & MAE & 0.149 & 0.137 & 0.155 & 0.125 & 0.141 & 0.132 & 0.133 & 0.133 & 0.148 & 0.137 & 0.139 & 0.009 \\ & $R^2$ & 0.710 & 0.671 & 0.672 & 0.779 & 0.706 & 0.724 & 0.745 & 0.758 & 0.655 & 0.656 & 0.708 & 0.044 \\ 
\bottomrule 
\end{tabular} 
\end{table*}

\begin{figure}
    \centering
    \includegraphics[width=0.9\textwidth]{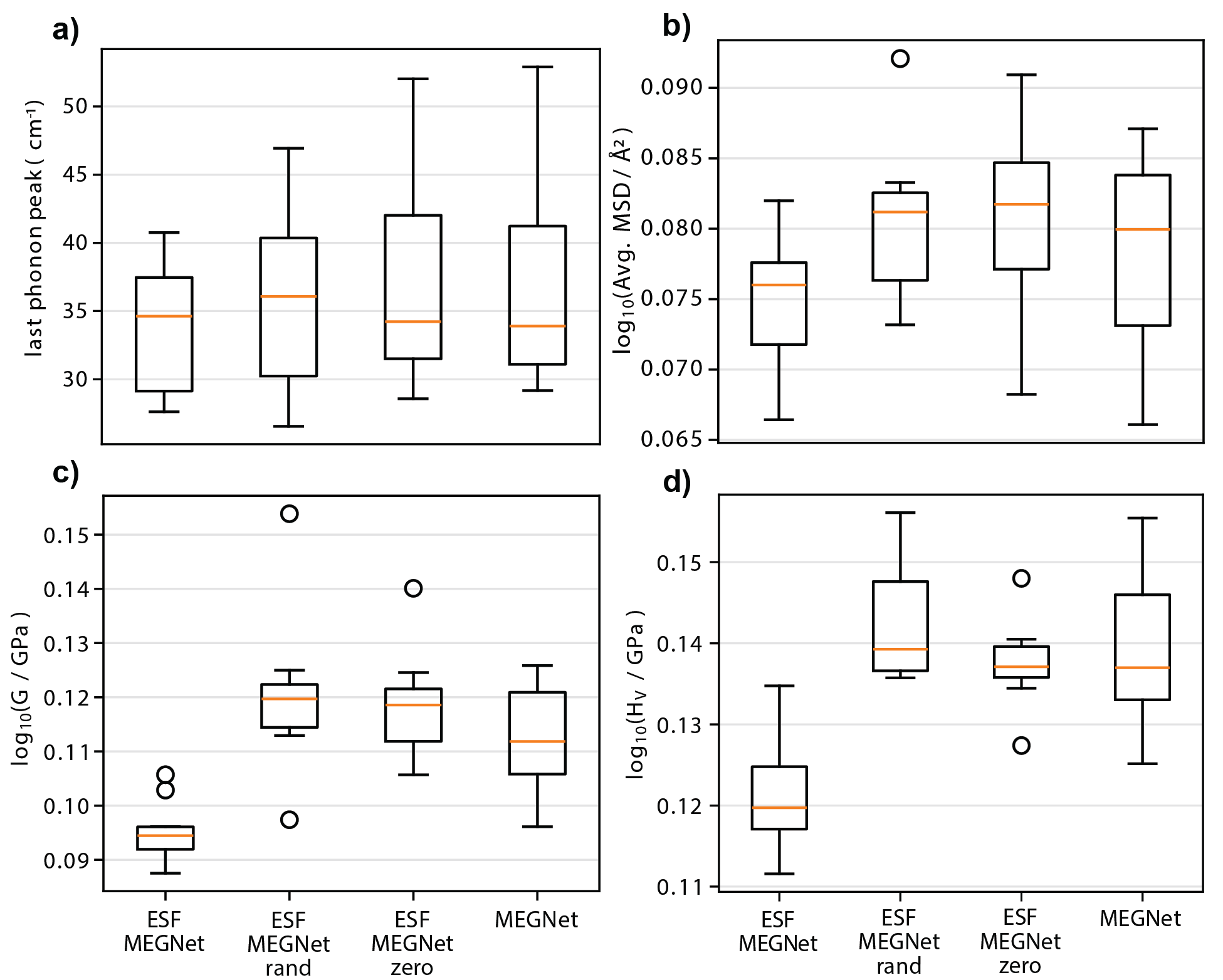}
    \caption{These box plots show the distribution of the MAEs across all ten folds from the four models for the following properties: a) last phonon DOS peak, b) Avg. MSD, c) G, and d) $H_V$. All plots show that ESF-MEGNet outperforms the baseline MEGNet and the two control models, ESF-MEGNet-zero and ESF-MEGNet-rand, which confirms that the quantum-chemical bonding descriptors add information to improve material property predictions.
    }
    \label{fig:SI_boxPlots}
\end{figure}


\end{document}